\documentclass[12pt]{article}

\usepackage{cite,bm}
\usepackage{epsfig}
\usepackage{amsmath,mathenv,slashed}
\usepackage{footmisc,amssymb}

\renewcommand{\thefootnote}{\fnsymbol{footnote}}

\newcommand\fverb{\setbox\fverbbox=\hbox\bgroup\verb}
\newcommand\fverbdo{\egroup\medskip\noindent%
			\fbox{\unhbox\fverbbox}\ }
\newcommand\fverbit{\egroup\item[\fbox{\unhbox\fverbbox}]}
\newbox\fverbbox

\newcommand{\rig}{\rightarrow}
\newcommand{\lrig}{\longrightarrow}
\renewcommand{\d}{{\mathrm{d}}}
\newcommand{\be}{\begin{eqnarray*}}
\newcommand{\ee}{\end{eqnarray*}}
\newcommand{\beq}{\begin{equation*}}
\newcommand{\eeq}{\end{equation*}}
\newcommand{\beeq}{\begin{equation}}
\newcommand{\eeeq}{\end{equation}}
\newcommand{\gl}[1]{(\ref{#1})}
\newcommand{\ta}[2]{ \frac{ {\mathrm{d}} #1 } {{\mathrm{d}} #2}}
\newcommand{\bee}{\begin{eqnarray}}
\newcommand{\eee}{\end{eqnarray}}

\def\lesssim{\mathrel{\raisebox{-.6ex}{$\stackrel{\textstyle<}{\sim}$}}}

\newcommand{\gev}{{\mathrm{~GeV}}}
\newcommand{\tev}{{\mathrm{~TeV}}}
\newcommand{\eps}{\varepsilon}
\newcommand{\ep}{\eps}
\newcommand{\alfas}{\alpha_s}
\newcommand{\jet}{{\rm{jet}}}

\begin{document}

\begin{titlepage} 
\nopagebreak  
{\flushleft{ 
        \begin{minipage}{6cm}
         FTUV--10--1004\\
         	KA-TP--30--2010\\
	SFB/CPP--10--88 \\ 
        	\end{minipage}        }}
\vfill 
\begin{center} 
{\LARGE \bf  
Precise predictions for (non-standard) \\[0.2cm] W$\bf{\gamma}$+jet production
} 
\end{center}
\vskip 1cm

\begin{center} {\large  \bf 
F.~Campanario$^{a,}$\footnote{{\tt{francam@particle.uni-karlsruhe.de}}},
C.~Englert$^{a,b,}$\footnote{\tt{c.englert@thphys.uni-heidelberg.de}}, 
and M.~Spannowsky$^{c,}$\footnote{{\tt{mspannow@uoregon.edu}}}
}
\end{center}

\vskip 0.2cm

\begin{center}   
\vskip .2cm  
$^{a}$
Institute for Theoretical Physics,
Karlsruhe Institute of Technology,\\
76128 Karlsruhe, Germany\\[0.3cm]
\noindent
$^{b}$ Institute for Theoretical Physics,
Heidelberg University,\\
69120 Heidelberg, Germany\\[0.3cm]
$^{c}$ Institute of Theoretical Science,
University of Oregon,\\
Eugene, OR 97403-5203, USA
\vskip 1.3cm     
\end{center} 
\nopagebreak 
\begin{abstract}
We report on a detailed investigation of the next-to-leading order (NLO) QCD corrections to 
$W\gamma$+jet production at the Tevatron and the LHC using a fully-flexible parton-level 
Monte Carlo program. We include the full leptonic decay of the $W$, taking into account 
all off-shell and finite width effects, as well as non-standard $WW\gamma$ couplings.
We find particularly sizable corrections for the currently allowed parameter range of
anomalous couplings imposed by LEP data. In total the NLO differential distributions reveal 
a substantial phase space dependence of the corrections, leaving considerable sensitivity 
to anomalous couplings beyond scale uncertainty at large momentum transfers in the 
anomalous vertex. 
\end{abstract} 
\vfill 
\end{titlepage} 
\newpage               

%
%
%
\setcounter{footnote}{0}
\section{Introduction}
Electroweak diboson production in association with a hard jet
is an important class of processes at hadron colliders such as
the Large Hadron Collider (LHC) or the Tevatron. Electroweak boson 
phenomenology generically provides a window to the electroweak symmetry breaking
sector, and diboson signatures with or without jets are therefore potentially sensitive to
new interactions beyond the Standard Model (BSM). Equally important, 
SM-diboson+jet production contributes to the irreducible background of new
physics searches in these channels. Hence, precise cross section 
predictions are mandatory to obtain a correct interpretation of possible excesses, 
which might be observed in the near future.
The total cross sections for the diboson+jet production processes are fairly large 
compared to the pure diboson production channels if extra jet emission at 
large available center-of-mass energy is kinematically unsuppressed.
The large one jet-inclusive cross section is mainly due to accessing 
the (anti)proton's gluon parton distribution function at small momentum fractions 
already at leading order (LO). 
At the same time, the gluon-induced partonic subprocesses opening up at $\mathcal{O}(\alfas)$ 
give rise to exceptionally large next-to-leading order 
(NLO) QCD corrections to inclusive diboson production, see Ref.~\cite{Ohnemus:1991kk}. 
Qualitatively similar observations have also been made for the NLO 
QCD corrections to various diboson+jet production processes in a series of 
recent publications \cite{Dittmaier:2007th,Dittmaier:2009un, 
Campbell:2007ev,Campanario:2009um,Binoth:2009wk,yettoapp,JiJuan:2010ga,
wzano}. 

In the present paper we extend our NLO calculation of $p\overline{p},pp\rig W^\pm\gamma+{\rm{jet}}+X$ 
\cite{Campanario:2009um} to anomalous $WW\gamma$ couplings\footnote{For 
convenience we refer to the computed processes as $W^\pm\gamma+{\rm{jet}}$ production 
even though we include all finite width and off-shell effects of the massive 
$W$.}. We review the SM $W^\pm\gamma+{\rm{jet}}$ phenomenology in detail and 
discuss its modifications due to anomalous couplings at NLO QCD precision. 
This allows us to discriminate between the effects of new physics in terms of effective interactions from the impact of higher order corrections.
Anomalous couplings searches represent benchmark tests for non-SM interactions at the LHC at small integrated 
luminosity ${\cal{L}}\lesssim 30~{\rm{fb}}^{-1}$ (see, e.g., Ref.~\cite{Dobbs:2005ev}). Measurement and discovery
strategies have received lots of attention, both from the theoretical (e.g. Refs.~\cite{Baur:1993ir,Dixon:1999di,Diakonos:1992qc,Chapon:2009hh}) and the experimental side 
(e.g. Refs.~\cite{Levin:2008zzb,Muller:2000,Dobbs:2005ev,Abazov:2009hk,Alcaraz:2006mx}). 
In this context, diboson production processes are important channels at the LHC because they exhibit large total rates,
and, in case of $W^\pm\gamma$ production, because they are sensitive to deviations of the underlying electroweak model 
from the SM via so-called radiation zeros.
These classical zeros of the amplitude in the $q \bar Q \rig W^\pm\gamma$ channels at the photon center-of-mass scattering angles 
$\cos\theta_\gamma=\mp 1/3$ are special to the completely destructive interference of gauge boson-radiation in an unbroken
renormalizable field theory, Ref.~\cite{Brown:1982xx}. Any deviation from QED by additional non-SM operators
ultimately destroys this characteristic radiation pattern. 
At the LHC, the antiquark direction is, in principle, indistinguishable from the quark direction 
because of the proton-proton initial state, and the radiation zero gets considerably washed out. 
"Signing" the quark direction according to the event's overall boost, which has been considered
in the context of dilepton asymmetries and electroweak mixing angle measurements 
\cite{Fisher:1994pw}, has been shown to efficiently lift the initial state's degeneracy 
in Ref.~\cite{Dobbs:2005ev}. Furthermore, the radiation zero remains present only if additional electromagnetically 
neutral (e.g. gluonic) radiation is collinear to the photon. Hence, additional QCD emission, as part of the NLO contribution 
to $W\gamma$ production, is dangerous to observing the radiation zero.
At the same time, the radiation zero becomes nearly impossible to measure in $W\gamma$+jet 
production \cite{Diakonos:1992qc}.

Crucial to significance-improving strategies \cite{Baur:1993ir} is therefore
an additionally-imposed jet veto\footnote{The jet veto also removes kinematical configurations which are less sensitive to anomalous couplings 
due to small momentum transfers in the $WW\gamma$ vertex from the total cross section, see below.}. 
Jet vetoing, however, is a delicate strategy in fixed-order perturbation theory from a theoretical point of view.
The observed reduction of scale dependence for the exclusive $W\gamma$ production at the LHC (see Ref.~\cite{Ohnemus:1991kk})
is predominantly due to excluding a region of phase space from the total inclusive cross section, which is well-accessible at the large 
available center-of-mass energy. This is also reflected in additional jet radiation becoming highly probable as part of the real emission contribution 
to the NLO diboson cross section, $\sigma (W\gamma +{\rm{jet}})/\sigma(W\gamma) \sim 3$. Hence, the dominant perturbative uncertainties 
result from the $W\gamma +{\rm{jet}}$ contribution, which is a leading order $\alfas$ contribution to NLO $W\gamma$ production.
Consequently, current Monte Carlo-driven strategies that involve jet vetos to measure anomalous couplings from
fits to high transverse momentum distributions (via e.g. neuronal net algorithms trained to the NLO $W\gamma$ distributions) 
inherit significant uncertainties, considerably larger than those given by scale variations of the exclusive NLO cross sections. 
By computing $W\gamma +{\rm{jet}}$ production at NLO accuracy, we are able
to realistically estimate the anomalous parameters' impact on the vetoed cross section and contribute a crucial
part towards modelling inclusive $W\gamma$ production at a higher perturbative precision.

We  organize this paper as follows:  Section~\ref{sec:calcdets} gives details on our Monte Carlo implementation 
and introduces the notion of anomalous $WW\gamma$ couplings to the reader. In Sec.~\ref{sec:numres}, we discuss the numerical results; we give total 
cross sections and differential distributions, both for SM and anomalous production. 
We also comment on the sensitivity to anomalous couplings in NLO QCD $W\gamma +{\rm{jet}}$ production at the LHC 
and we quote $W\gamma +{\rm{jet}}$ cross sections for selection cuts adapted to anomalous couplings' searches.
The phenomenological impact of anomalous couplings on $W\gamma$+jet production at the Tevatron is too small
for parameter choices that are compatible with the bounds imposed by LEP data, Ref.~\cite{Alcaraz:2006mx}. 
Hence, we only quote Tevatron results for SM-like production in Sec.~\ref{sec:tevares}.
Section~\ref{sec:summary} closes with a summary and gives an outlook to future work.
Our Monte Carlo code will become publicly 
available with an upcoming update of {\sc Vbfnlo} \cite{Arnold:2008rz}.

\section{Details of the calculation}
\label{sec:calcdets}
There are three contributing partonic subprocesses at $\mathcal{O}(\alpha^3\alfas)$ for $p\overline p, pp\rig e^-\bar \nu_e \gamma j+ X$,
\begin{subequations}
\label{wgammajloprocs}
\bee[lcl]
\label{wgammajloprocsa}
q \bar Q &\lrig& \ell^- \bar \nu_\ell \gamma  g \,,\\
\label{wgammajloprocsb}
\bar Q g&\lrig& \ell^- \bar \nu_\ell \gamma  \bar q \,, \\
\label{wgammajloprocsc}
q g &\lrig& \ell^-\bar \nu_\ell \gamma   Q \,,
\eee
\end{subequations}
not counting the subprocesses, which follow from interchanging the beam directions. 
We use the shorthand notation $q=(d,s)$ and $Q=(u,c)$ and assume a diagonal CKM matrix.
At the LHC, a non-diagonal CKM matrix decreases our leading order results only at the per mil-level. 
Unitarity of the (non-diagonal) CKM matrix guarantees that all CKM-dependence drops out for flavor-blind 
observables computed from the dominant gluon-induced subprocesses.
At the Tevatron, we find our cross sections decreased by about 3\%. 
Both modifications are smaller than the residual scale dependence at NLO, so that a diagonal CKM matrix is an adequate
approximation for our purposes. Bottom quark contributions are absent at LO for the above approximations and can 
be further suppressed experimentally by $b$ vetoing, and we therefore neglect bottom contributions throughout the
computation.

\begin{figure}[!t]
\begin{center}
\parbox{0.5\textwidth}{
\epsfig{file=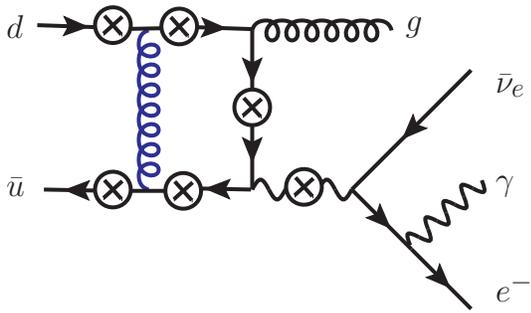, scale=0.5}}
\parbox{0.44\textwidth}{
\caption{\small \label{fig:feynvirt}  Representative Feynman graph contributing to the virtual corrections to
the partonic subprocess $\bar u d\rightarrow  e^-\bar\nu_e \gamma g$ at
${\cal{O}}(\alpha^3\alpha_s^2)$. 
The crosses mark points where the photon can be attached to the quarks or
the $W$ for the fixed gluon-$W$ order along the quark line.}}
\end{center}
\end{figure}

The LO matrix elements of Eq.~\gl{wgammajloprocs} are calculated using {\sc Helas} routines  \cite{Murayama:1992gi} 
generated with {\sc MadGraph} \cite{Alwall:2007st}. Although we refer to the processes as $W^\pm \gamma$+jet production for
convenience, we include all off-shell and finite width effects of the $W$'s decay to leptons, as well as the photon's coupling to the charged
final state lepton and calculate the full QCD corrections to the processes $pp,p\overline p \rig \ell^-\bar\nu_\ell \gamma j + X$ and 
 $pp, p\overline p\rig \ell^+\nu_\ell \gamma j + X$ at ${\cal{O}}(\alpha^3\alpha_s^2)$. Representative virtual Feynman graph topologies are sketched in 
 Fig.~\ref{fig:feynvirt}.
We perform the numerical phase space integration with a modified version of \textsc{Vegas} \cite{Lepage:1977sw}, 
which is part of the \textsc{Vbfnlo} package. 
We divided up the integration and explicitly sum over different channels that are optimized for the two and 
three-body decay of the (off-shell) $W$ boson, $pp, p\overline p\rig W\gamma j \rig \ell\nu\gamma j$ and 
$pp,p\overline p\rig W j \rig \ell\nu\gamma j$, respectively. For additional details on the phase space integration's 
validation against \textsc{MadEvent} \cite{Alwall:2007st} and \textsc{Sherpa} \cite{Gleisberg:2008ta}, 
we refer the reader to our previous publication \cite{Campanario:2009um}.
The numerical implementation includes the finite width of the $W$ by
using the fixed width scheme\footnote{This is also the width scheme used by \textsc{MadGraph}.} 
of Ref.~\cite{Denner:1999gp}: The weak mixing angle is taken to be real and we use
Breit-Wigner propagators for the massive $W$ throughout.

%
%

The counter term-renormalized virtual amplitude for, e.g., $\bar u d\rig \ell^-\bar\nu_e\gamma g$ (Fig.~\ref{fig:feynvirt}) in conventional 
dimensional regularization $d=4-2\eps$ can be cast into 
the form,
\beeq
\begin{split}
\label{eq:virtboxline}
{\cal M}^{\rm{1-Loop}}_{\rm{Virt+CT}}&=
 \frac{\alpha_s(\mu_{R}^2)}{4 \pi} 
{ \mathcal{M}_{{\rm{LO}}} \over \Gamma(1-\eps)}
\left[ \frac{1}{2} 
\left\{ \left (\frac{4 \pi \mu_{R}^2}{-u }\right)^{\eps} +
\left (\frac{4 \pi \mu_{R}^2}{-t}\right)^{\eps} \right\}
\left(-\frac{C_A}{\eps^2}-\frac{\gamma_{g}}{\eps}\right) \right. \\
&+ \frac{C_A}{2C_F} \left \{ \left( \frac{4 \pi \mu_{R}^2}{-u} 
\right)^{\eps} +  
\left( \frac{4 \pi \mu_{R}^2}{-t} \right)^{\eps}-2  
\left( \frac{4 \pi \mu_{R}^2}{-s} \right)^{\eps} \right\}
\left(-\frac{C_F}{\eps^2}-\frac{\gamma_q}{\eps}\right) \\
&+
2 \left (\frac{4 \pi \mu_{R}^2}{-s} \right)^{\eps}
 \left(-\frac{C_F}{\eps^2} -\frac{\gamma_q}{\eps}\right)
\bigg] + \widetilde{\cal{M}}_{\rm{Virt}}(-s,-t,-u) \, ,
\end{split}
\eeeq
with $C_F=4/3$ and $C_A=3$ denoting the casimirs of the fundamental and adjoint representations, respectively.
The virtual amplitude exhibits an identical structure in color space as the Born matrix element and we implicitly
assume the ${\mathfrak{su}}(3)$ generator in the fundamental representation to be part of the definitions of $\cal{M_{\rm{LO}}}$
and $\widetilde{\cal{M}}_{\rm{Virt}}$.
The constants $\gamma_i$ are fixed by the $SU(3)$ representations of the $N_F=5$ active quark flavors and the gluons,
\bee
\gamma_q = \gamma_{\bar q} = \frac{3}{2} C_F \,, \qquad
\gamma_{g} = \frac{11}{6} C_A - \frac{2}{3} T_r N_F \,,
\eee
where $T_r=1/2$ is the Dynkin index of the fundamental representation.
$s,t,u$ denote the familiar Mandelstam variables of a $2\rig 2$ process, taken to be space-like.
Depending on the subprocess, the analytical continuation to physical kinematics is performed automatically
within our numerical code by effectively restoring the propagators' $\ep$ description.
To arrive at the correct logarithms when evaluating Eq.~\gl{eq:virtboxline} for physical $s,t,u$, we have
to write, e.g.,
\bee
\label{eq:virtancont}
\log\left(-s \right ) = \log ( |s| ) -i\pi\,\Theta(-s)
\eee
for the $q\bar Q$ induced subprocesses for which $s$ is a time-like quantity. Similar formulae have to be taken into account for the 
analytical continuation of the dilogarithms and can be inferred from the literature, e.g. from Ref. \cite{vanHameren:2005ed}, appendix C.
$\widetilde{\cal{M}}_{\rm{Virt}}$ in Eq.~\gl{eq:virtboxline} represents finite contributions that embrace tensor coefficients and 
fermion chains after 
algebraic manipulations. We also implicitly include the so-called rational terms to our definition of $\widetilde{\cal{M}}_{\rm{Virt}}$. These 
finite contributions arise from the interplay of $d$-dimensional numerator and denominator algebra in the limit $\eps\rig 0$ 
(see,~e.g.,~\cite{Bredenstein:2008zb}). To compute the tensor coefficients, we apply the Pas\-sa\-ri\-no-Veltman recursion 
up to box topologies \cite{Passarino:1978jh} and the Denner-Dittmaier reduction for pentagon graphs \cite{Denner:2002ii}.

The full virtual amplitude can be assembled from elementary building blocks, that divide the one loop graphs
into certain groups. This strategy has already been applied to calculate a series of different processes 
at NLO QCD precision in Ref.~\cite{Englert:2008wp}, both in the SM and beyond. Concretely, we sum all self-energy, triangle, box and pentagon 
corrections to a quark line with three attached gauge bosons of a given order to yield a single numerical routine. A similar routine
is constructed from a quark line with two attached gauge bosons (see Fig.~\ref{fig:feynvirt}).
These routines are set up with an in-house framework that partly uses {\sc{FeynArts}} \cite{Hahn:2000kx} and 
{\sc{FeynCalc}} \cite{Mertig:1990an}. From these building blocks we construct the full SM loop amplitude for a given 
subprocess by trivial permutations of the bosons' momenta and polarization vectors, which also encode the two and three-body decay of the
massive $W$ depending on the building block. Replacing the SM polarization vectors by polarization vectors modified by including anomalous $WW\gamma$ couplings, 
we straightforwardly generalize our NLO calculation to anomalous $W \gamma+\rm{jet}$ production. 
The renormalization in Eq.~\gl{eq:virtboxline} is performed on-shell for the wave functions and we use the $\overline{\rm{MS}}$-scheme
to renormalize the strong coupling constant.
The virtual corrections to all other subprocesses of Eq.~\gl{wgammajloprocs} can be recovered 
from Eq.~\gl{eq:virtboxline} by crossing and analytical continuation analogous to Eq.~\gl{eq:virtancont}.
%
%
\begin{figure}[t!]
\begin{center}
\parbox{0.50\textwidth}{
\epsfig{file=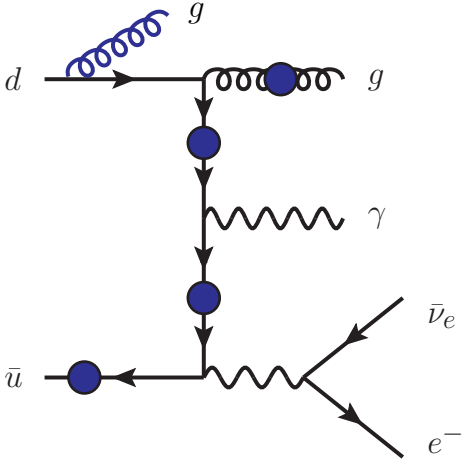, scale=0.50}}
\parbox{0.44\textwidth}{
\caption{\small
\label{fig:feynrem}
Sample Feynman graph contributing to the partonic real emission subprocess
$\bar ud\rightarrow e^- \bar\nu_e \gamma g g$ at ${\cal{O}}(\alpha^3\alpha_s^2)$. 
The gluon can also be attached to the
quark and gluon lines at the positions marked by the circles. 
Feynman graph topologies, where the photon is radiated off at different positions 
analogous to Fig.~1, are not shown.
}}
\vspace{-0.3cm}
\end{center}
\end{figure} 

To speed up the numerical evaluation of the 90 subprocesses of the real emission contribution, we compute the respective
matrix elements using optimized code that employs the spinor helicity formalism of Ref.~\cite{Hagiwara:1988pp}. The generalization to matrix elements with 
anomalous couplings is again performed by modifying the three-body effective $W$ polarization vector to include the
non-SM interactions of Sec.~\ref{subsec:anom}. The subprocesses for $pp,p\overline p \rig e^-\bar\nu_e\gamma j j+X$ 
at ${\mathcal{O}}(\alpha^3\alfas^2)$ can be classified, modulo crossing, flavor summation, and initial state interchange, into
\begin{subequations}
\label{eq:remsubprocs}
\bee[lclqqlcl]
\bar{u} u &\longrightarrow &\ell^- \bar{\nu}_\ell \gamma \bar{d} u \,, &
\bar{u} u  &\longrightarrow &\ell^- \bar{\nu}_\ell \gamma \bar{s} c  \,, \\
\bar{d} d &\longrightarrow &\ell^- \bar{\nu}_\ell \gamma \bar{u} d \,, &
\bar{d} d &\longrightarrow& \ell^- \bar{\nu}_\ell \gamma \bar{c} s \,, \\
\bar{u} d &\longrightarrow &\ell^- \bar{\nu}_\ell \gamma  g g \,.
\eee
\end{subequations}
Representative Feynman topologies for the last line's subprocess are indicated in Fig.~\ref{fig:feynrem}. 
We store intermediate numerical results common to all subprocesses 
and reuse them whenever possible to speed up the numerical implementation.
All matrix elements have been checked explicitly against code generated with {\sc MadGraph} as well as against {\textsc{Sherpa}} for 
integrated cross sections (cf. \cite{Campanario:2009um}). Applying the dipole subtraction of Catani and Seymour \cite{Catani:1996vz}, 
we have verified our implementation against code generated with the \textsc{MadDipoles} \cite{Frederix:2008hu} package. 
Our code is optimized such that intermediate dipole results are reused in order to 
avoid redundant dipole or Born-level matrix element calculations. We also recycle the dipoles' Born-level matrix elements into
the integration of the finite collinear remainder, which is left after renormalization of the parton distribution functions 
\cite{Altarelli:1977zs, Catani:1996vz}. We integrate this contribution over the real emission phase space
applying the phase space mappings of Ref.~\cite{Figy:2003nv}.
The remaining IR-singularities of the virtual matrix element Eq.~\gl{eq:virtboxline} cancel analytically
against the one-parton phase space-integrated dipoles, symbolically denoted by $\left\langle{\bm I} \right\rangle$ in the language of 
Ref.~\cite{Catani:1996vz}. For, e.g., the $\bar qQ$-induced channels adding the $\left\langle{\bm I} \right\rangle$ operator yields
\begin{multline}
\label{virt:finiterem}
2\,{\rm{Re}}\left( {\cal M}^{{\rm{1-Loop,}}q\bar Q \rig g}_{\rm{Virt+CT}}  \left[{\cal{M}}^{q\bar Q\rig g}_{\rm{LO}}\right]^\ast \right)+
\left\langle{\bm I} \right\rangle
 = \frac{\alfas(\mu_R^2) }{2\pi}  
\frac{ |{\cal M}^{\bar q Q \rig g}_{\rm{LO}}|^2}{\Gamma (1-\eps )}
\\
\times
\bigg[{1\over 2}\left\{\left(\frac{4\pi\mu_R^2}{-t}\right)^{\eps}+
\left(\frac{4\pi\mu_R^2}{-u}\right)^{\ep}\right\}
\left( K_g+\gamma_g-\frac{\pi ^2 C_A}{3}\right) 
-\left(K_q+\gamma_q
-\frac{\pi ^2 C_F}{3} \right)
\\
\times\left\{ {C_A-2C_F\over C_F} 
\left(\frac{4\pi\mu_R^2}{s}\right)^{\eps}-{C_A\over 2C_F}
\left(\frac{4\pi\mu_R^2}{-t}\right)^{\ep}
-{C_A\over 2C_F}\left(\frac{4\pi\mu_R^2}{-u}\right)^{\ep}
\right\}\, \bigg] \\
+ 2\,{\rm{Re}}\left(\widetilde{\cal{M}}'_{\rm{Virt}} \left[{\cal{M}}^{q\bar Q\rig g}_{\rm{LO}}\right]^\ast\right)\,,
\end{multline}
with
\beeq
\label{kcapp}
K_q = K_{\bar q} = \left( \frac{7}{2} - \frac{\pi^2}{6} \right) C_F \,, \qquad
K_{g} = \left( \frac{67}{18} - \frac{\pi^2}{6} \right) C_A - \frac{10}{9}
T_r N_F \,,
\eeeq
where we have already performed the analytical continuation to time-like $s$.
The primed $\widetilde{\cal{M}}'_{\rm{Virt}}$ in Eq.~\gl{virt:finiterem} indicates that we deal with a changed 
finite piece compared to Eq.~\gl{eq:virtboxline} due to the analytic continuation. Note that we again include
the Born level color structure into the definitions of the amplitudes; summing and averaging over colors and 
spins yields trivial subprocess-dependent additional prefactors. 
Equation \gl{virt:finiterem} is explicitly free of IR divergencies and hence finite for $\ep\rig 0$.
A similar factorization and cancellation up to box diagrams has been demonstrated in Ref.~\cite{Figy:2007kv}.

\subsection{Anomalous WW$\bf{\gamma}$ couplings}
\label{subsec:anom}
We parametrize deviations of the SM electroweak sector by extending the SM Lagrangian by the most general,
QED-preserving, Lorentz and $\mathcal{CP}$-invariant operators up to dimension six \cite{Hagiwara:1986vm},
\beeq
\label{anovertex}
\mathcal{L}_{WW\gamma} =
-ie \left[ W_{\mu\nu}^\dagger W^\mu A^\nu- W_\mu^\dagger A_\nu W^{\mu\nu}
+\kappa  W_\mu^\dagger W_\nu F^{\mu\nu} +{\lambda\over m_W^2} W_{\lambda\mu}^\dagger W^\mu_\nu
F^{\nu\lambda}\right]\,.
\eeeq
It is customary to express the $\kappa$-induced deviation from the SM operators\footnote{We recover the electroweak part 
of the SM by choosing the parameters $\kappa=1$ and $\lambda=0$.} by $\Delta\kappa=\kappa-1$. 
The parameters $\Delta\kappa$ and $\lambda$ are related to the electric quadrupole 
moment $Q_W$ and magnetic dipole moment $\mu_W$ of the $W$ boson \cite{Hagiwara:1986vm},
\bee
\mu_W={e\over 2m_W} (2+\Delta\kappa +\lambda)\,,\qquad
Q_W=-{e\over m_W^2} (1+\Delta\kappa-\lambda)\,.
\eee

Retaining unitarity at high energies is crucial to meaningfully modeling
physics beyond the SM. On the one hand, if probability conservation is violated, the cross section receives a sizable
contribution from probing the matrix elements at large invariant masses, even though the parton luminosities tame 
the matrix elements' unphysical growth in this particular phase space region. 
On the other hand, if unitarity is conserved, the phenomenology is mainly dominated by comparatively low invariant masses
by the same reason. In order not to violate unitarity, the parameters $\Delta\kappa$ and $\lambda$ have to
be understood as low-energy form factors, and their precise momentum dependence does depend sensitively on physics beyond the SM. 
However, a widely-used phenomenological parametrization is (cf. Refs.~\cite{Baur:1993ir,Hagiwara:1986vm}),
\bee
\label{eq:anomform}
\Delta\kappa = {\Delta\kappa_0 \over \big(1+(p_\gamma + p_W)^2/\Lambda^2\big)^{n_\kappa}}\, ,
\qquad
\lambda = {\lambda_0 \over \big(1+(p_\gamma + p_W)^2/\Lambda^2\big)^{n_\lambda}}\, ,
\eee
where $\Lambda$ represents the scale at which the beyond-the-SM interactions become strong, i.e. the scale at which $W$ compositeness
is resolved. $p_\gamma$ and $p_W$ denote the final state momenta of the photon and the $W$, respectively.
Unitarity imposes $n_\kappa> 1/2$ and $n_\lambda>1$ \cite{Baur:1987mt}; customary choices by experimentalists 
are dipole profiles $n_\kappa=n_\lambda=2$ \cite{Levin:2008zzb,Muller:2000,Dobbs:2005ev,Abazov:2009hk,Alcaraz:2006mx}.
Note that we do not include anomalous ${\cal{CP}}$-violating operators since they have already been tightly constrained
by measurements of the neutron electric dipole moment \cite{Amsler:2008zzb,Baur:1993ir}.

We include the anomalous interactions by constructing purpose-built {\sc{Helas}} routines that implement the
effective Lagrangian of Eq.~\gl{anovertex} in a straightforward way using \textsc{FeynRules} (Ref.~\cite{Christensen:2008py}) 
to arrive at the analytical expression for the anomalous vertex function.

\section{Numerical results}
\label{sec:numres}
\subsection{Selection criteria, general Monte Carlo input and photon isolation}
Throughout, we use CTEQ6M parton distributions \cite{Pumplin:2002vw} with $\alpha_s(m_Z)=0.118$ at NLO, and the CTEQ6L1 set at LO. 
We choose $m_{Z}=91.188~\rm{GeV}$, $m_{W}=80.419~\rm{GeV}$ and $G_F=1.16639\times 10^{-5}~\textnormal{GeV}^{-2}$ 
as electroweak input parameters and derive the electromagnetic coupling $\alpha$ and the weak mixing angle from SM-tree level relations. 
The center-of-mass energy is fixed to $14~\rm{TeV}$ for LHC and to $1.96\tev$ for Tevatron collisions. We consider one family of light leptons 
in the final state, which we treat as massless, i.e. we quote results for $pp,p\overline p \rig e^-\bar \nu_e\gamma j + X$ \underline{or} 
$pp,p\overline p \rig \mu^-\bar \mu_e\gamma j + X$ when we speak of $W^-\gamma +{\rm{jet}}$ production.

Jets are recombined from partons with pseudorapidity $|\eta|\leq 5$ applying the algorithm of Ref.~\cite{Catani:1993hr} with resolution parameter $D=0.7$. 
The reconstructed jets are required to lie in the rapidity range 
\bee 
|y_j | \leq 4.5\,.
\eee
The charged lepton and the photon are required to fall into the rapidity coverage of the electromagnetic
calorimeter, i.e. we impose
\bee
|\eta_\ell|\leq 2.5\,, \qquad |\eta_\gamma|\leq 2.5\,.
\eee
In order not to spoil the cancellation of the IR singularities and to minimize contributions from non-perturbative jet fragmentation associated with 
collinear photon-jet configurations, we apply the photon isolation criterion of Ref.~\cite{Frixione:1998jh}:
\beeq
\label{eq:photonisolation} 
\sum_{i, R_{i\gamma}<R} p_T^{{\rm{parton}}, i} \leq \frac{1- \cos R}{1- \cos \delta_0}\,  p_T^\gamma \quad \forall R\leq \delta_0\,.
\eeeq
The index $i$ in Eq.~\gl{eq:photonisolation} runs over all partons found in a cone around the photon of size $R$ in the azimuthal 
angle--pseudorapidity plane. The IR-safe cone size around the photon is given by $\delta_0$, and, in principle, $p_T^\gamma$ can be 
replaced by an arbitrary energy scale $\mathcal{E}$, which then determines the penetrability of the photon cone by soft QCD radiation.

\begin{figure}[t]
\centering
\includegraphics[scale=0.92]{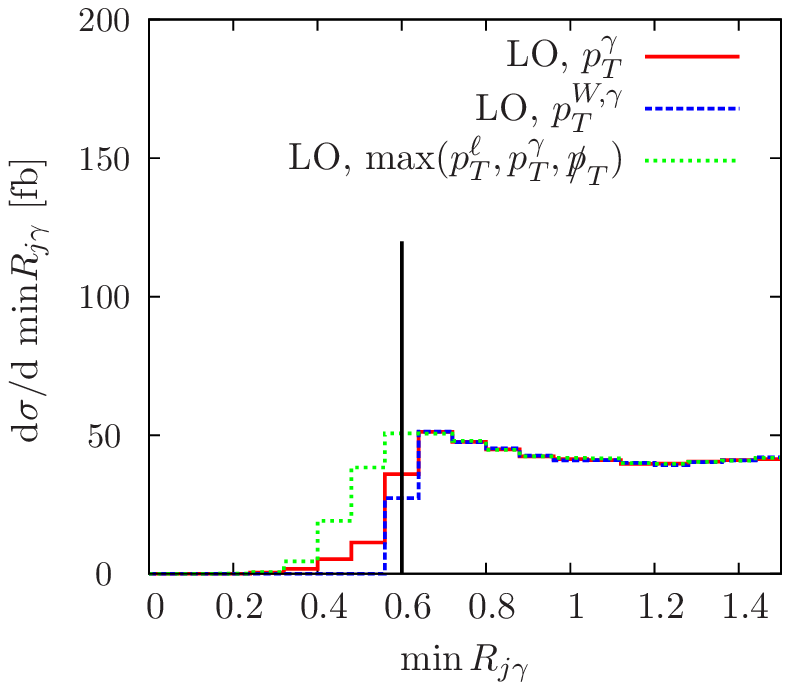}
\hfill
\includegraphics[scale=0.92]{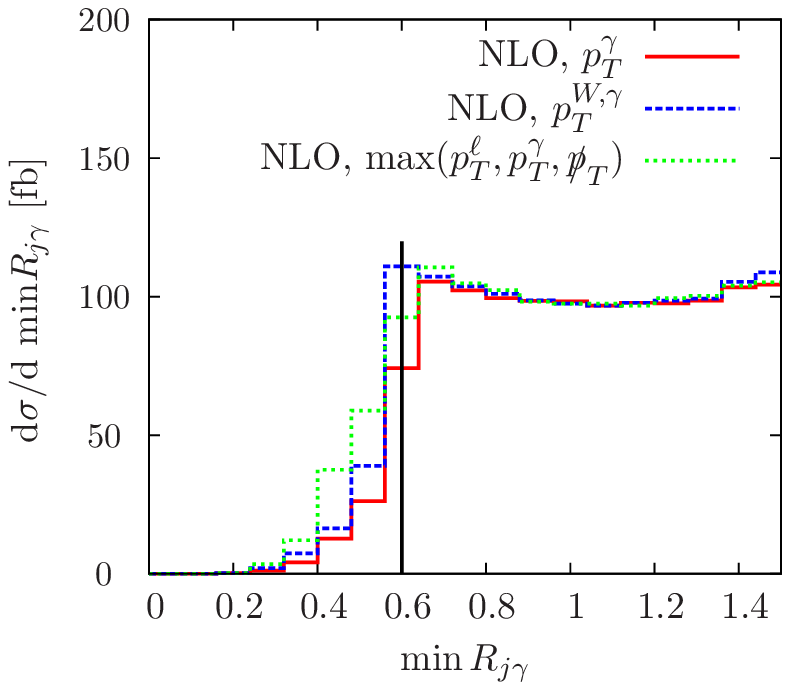}\\[0.3cm]
\includegraphics[scale=0.92]{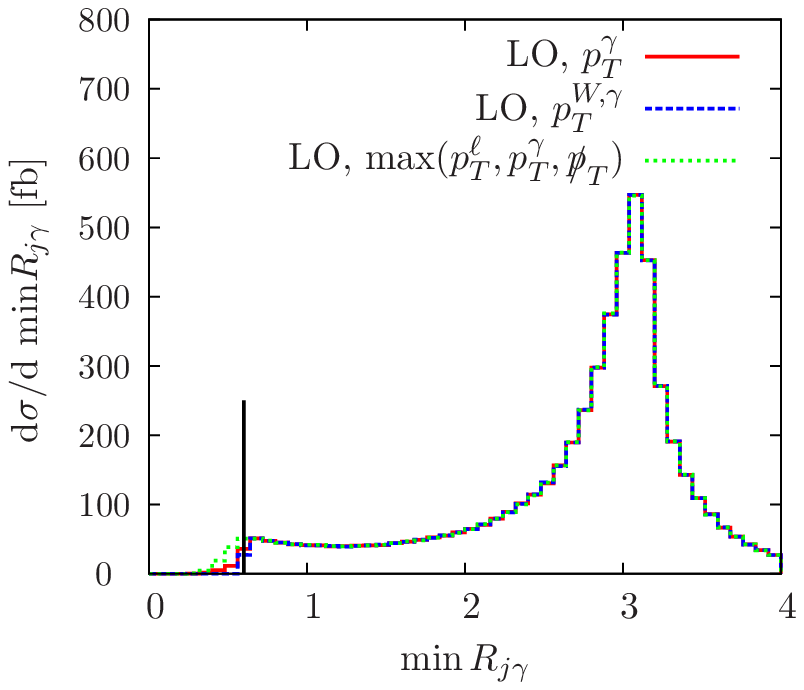}
\hfill
\includegraphics[scale=0.92]{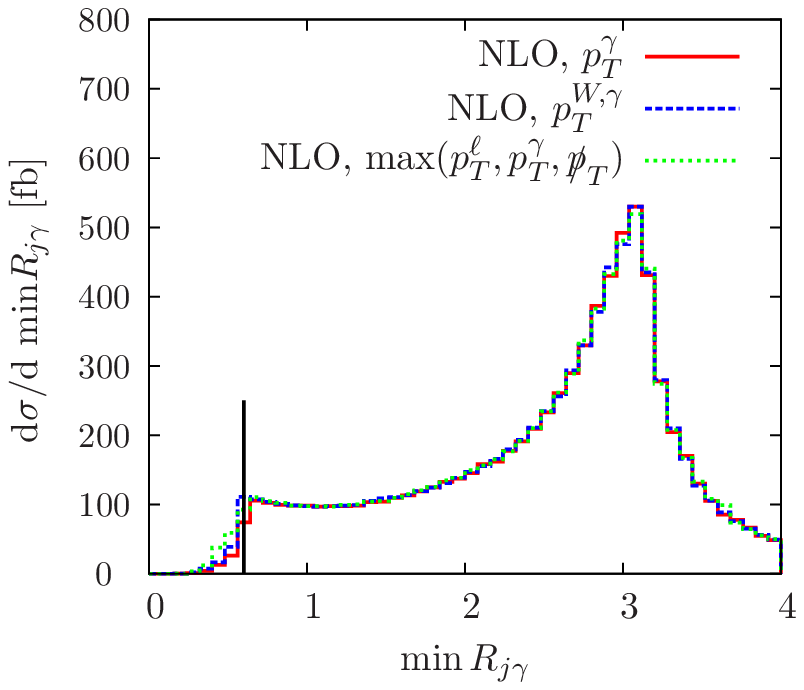}
\caption{\small \label{fig:LHCwgammaisolcls} Minimal photon-jet separation in the azimuthal angle -- pseudorapidity plane 
for $W^-\gamma$+jet production at the LHC. Plotted are the leading order and next-to-leading order distributions 
for different isolation scales ${\cal{E}}=p_T^\gamma$, $p_T^{W,\gamma}$, $\max(p_T^\ell,p_t^\gamma,\slashed{p}_T)$ 
in Eq.~\gl{eq:photonisolation}. The upper row displays the lower row's $\min R_{j\gamma}$ distributions 
around the isolation cone in more detail. The isolation parameter is $\delta_0 = 0.6$ (indicated by the vertical line), 
and $\mu_F=\mu_R=100\gev$. Cuts are chosen as described in the text.
\vspace{-0.3cm}}
\end{figure}

In a realistic experimental setting, the photon identification efficiency depends on the number of finite-sized calorimeter towers 
that enter the candidate photon reconstruction (cf. Ref.~\cite{dc1}). It is therefore worthwhile to note that the isolation scale choice
implicitly enters in the photon definition and different scale choices will be accompanied by different identification efficiencies for 
small values of $\delta_0$. A detailed investigation of this direction should include effects ranging from pile-up and underlying 
event to NLO-corrected jet fragmentation and is beyond the scope of this work.
For values of $\delta_0$ much larger than the electromagnetic calorimeter cell size of $\Delta R \approx 0.04$, however, 
Eq.~\gl{eq:photonisolation} is a ``sliding-cut'' prescription, which is experimentally well-defined on the level of already reconstructed particles.
We can therefore compare the impact of the IR-safe isolation criterion to the NLO 
scale uncertainty, which turns out to be of order 10\%. Replacing $p_T^\gamma$ in Eq.~\gl{eq:photonisolation} 
by other intrinsic scales to the process, e.g. by $\max p_T^j$, corrects our NLO results at the level of $1\%$ for inclusive cuts 
and $\delta_0=0.6$ (Fig.~\ref{fig:LHCwgammaisolcls}).
Comparing the LO and NLO inclusive distributions we find a large net increase of the differential cross section.
The differences in the choice of the isolation scale, however, are only
visible in the jet-photon separation distribution around the photon cone $\min R_{j\gamma} \approx \delta_0$.
For larger separations $\min R_{j\gamma} > \delta_0$ we do not find any notable phenomenological impact of the isolation scale once we take into
account the cross section's residual scale dependence on $\mu_R,\mu_F$.
For the isolation scale ${\cal{E}}=p_T^{W,\gamma}=| \bm{p}_T^\gamma + \bm{p}_T^\ell + \slashed{\bm{p}}_T |$ the 
threshold behavior around $\delta_0$ changes most significantly when comparing LO and NLO distributions 
(we denote the neutrino's four-momentum by $\slashed{p}$ in the following). The behaviour at LO is a consequence of $p^{W,\gamma}_T=p_T^j$ 
and the $W$ recoiling against the photon-jet pair if the jet is emitted around the photon cone. 
Therefore, probing smaller parton-photon separations effectively means increasing 
the transverse momentum of the $W$ for central events. The isolation scale, however, is set by the jet itself. 
Due to the exponential drop-off of the $p_T^W = |\bm{p}_T^\ell + \slashed{\bm{p}}_T | $ spectrum, collinear jet-photon configurations 
are highly attenuated for separations smaller than $\delta_0$ at LO, see Fig. \ref{fig:LHCwgammaisolcls}.
At NLO the kinematical LO correlation of the jet and $W\gamma$-system 
is modified by additional parton emission, which allows the $W$ to be emitted at smaller transverse momenta. 
Thus, QCD radiation into the photon cone around the threshold $\delta_0$ becomes more probable than at LO. 
At the same time, however, $p_T^{W,\gamma}$ decreases and more partons get vetoed at distances
smaller than $\delta_0$, and a steep drop-off is still visible in $\min R_{j\gamma}$ around $\delta_0$ at NLO for 
${\cal{E}}=p_T^{W,\gamma}$ in Fig.~\ref{fig:LHCwgammaisolcls}. In addition, we add events with a positive-definite weight to 
the minimum-separation distribution with the second resolved jet approximately balancing the jet-photon-$W$ system in $p_T$,
which also modifies the $\min R_{j\gamma}$ threshold behaviour.

We now move on to investigate the general features of $W\gamma + {\rm jet}$ production with inclusive cuts on jets, photon, and lepton; we require
\beeq
\label{eq:ptcut}
p_T^j\geq 50\gev,\quad p_T^\ell\geq 20\gev, \quad  p_T^\gamma\geq 20\gev\,.
\eeeq
To avoid the collinear photon-lepton configurations we impose a finite separation in the azimuthal angle-pseudorapidity plane of
\beeq
\label{eq:lgammacut}
R_{\ell\gamma} = \sqrt{ \Delta\phi_{\ell\gamma}^2 + \Delta\eta_{\ell\gamma}^2} \geq 0.2\,.
\eeeq
For the jet-lepton separation we choose
\beeq
R_{  j\ell}  \geq 0.2\,,
\eeeq
%
%
\begin{table}[!b]
\begin{center}
\begin{tabular}{c c c}
\hline
& 
$\sigma^{\rm{NLO}}_{\rm{incl}}$ [fb]& 
$K=\sigma^{\rm{NLO}}_{\rm{incl}}/\sigma^{\rm{LO}}$ \\
\hline
$W^-\gamma j$  & $615.3 \pm 2.8$   &  $1.49$ \\
$W^+\gamma j$ & $736.5 \pm 3.5$   &  $1.41$ \\ 
\hline
\end{tabular}
\caption{\small\label{tab:kfactorswga} Inclusive next-to-leading order cross sections and total $K$ factors for the processes $pp\rightarrow e^+\nu_e\gamma j+X$ and $pp\rightarrow e^-\bar\nu_e\gamma j+X$ at the LHC for identified renormalization and factorization scales, $\mu_R=\mu_F=100~\rm{GeV}$. The cuts are chosen as described in the text.}
\end{center}
\end{table}
\hspace{-0.15cm}and for the photon isolation we impose
\beeq
\delta_0=0.6\,.
\eeeq
It is customary to also analyze the cross sections' behavior with an additional 'no resolvable $2^{\rm{nd}}$ jet'--criterion \cite{Dittmaier:2007th}, i.e. a veto on the second jet, if it gets
resolved,
\bee[lqqTl]
\label{eq:wajveto}
 p_T^{j,\rm{veto}}\geq 50\gev\,, \quad |y^{\rm{veto}}_j|\leq 4.5\, & (exclusive NLO)\,.
\eee
This allows us to identify the dominant contributions to the NLO-inclusive cross section. In addition,
it was shown in Ref.~\cite{Campanario:2009um} that this cut leads to seemingly stable exclusive $W\gamma$+jet production cross sections at the LHC.
Similar observations have been made for the other diboson+jet cross sections provided in Refs.~\cite{Dittmaier:2007th,Dittmaier:2009un,Campbell:2007ev,
Binoth:2009wk,yettoapp,JiJuan:2010ga}. The stabilization of the exclusive cross section does, however, not provide a reliable estimate of the perturbative 
cross sections' uncertainties over the whole phase space. We will discuss this in more detail in Sec.~\ref{sec:lhc}.

\subsection{NLO QCD W${\bf{\gamma}}$+jet production in the SM at the LHC}
\label{sec:lhc}
%
%
\begin{figure}[t]
\begin{center}
\epsfig{file=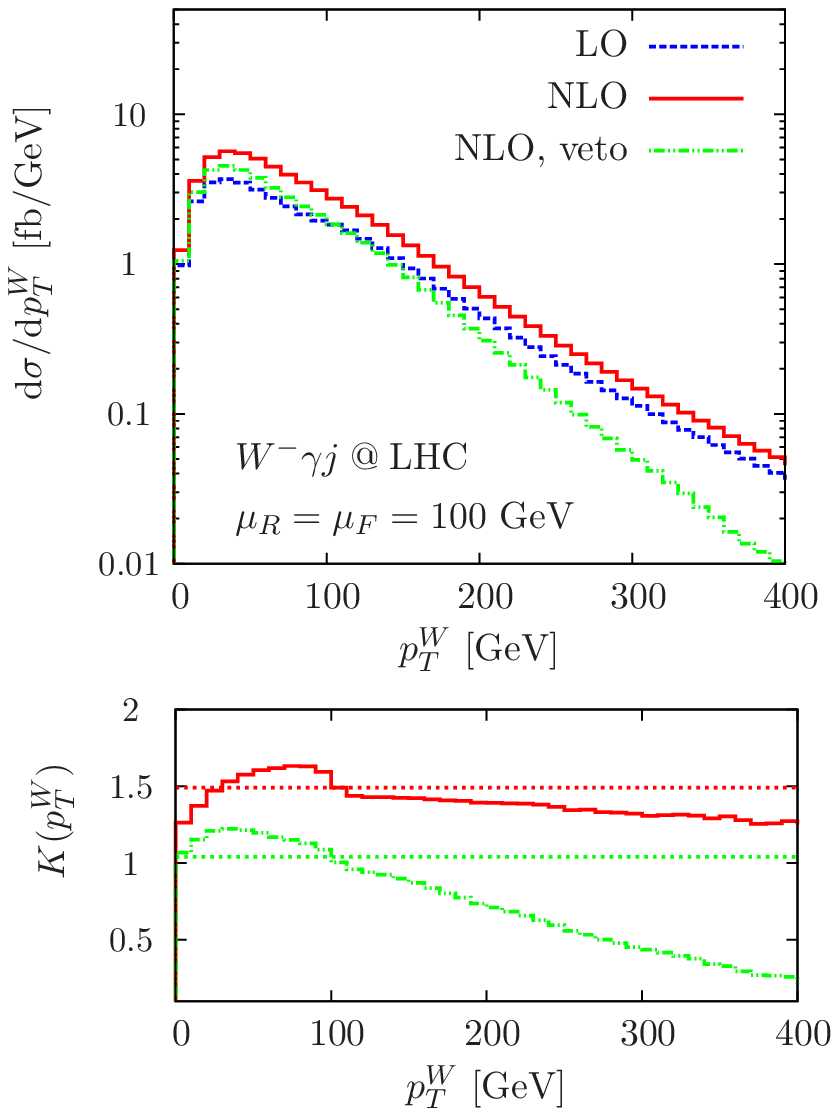, width=0.47\textwidth}
\hfill
\epsfig{file=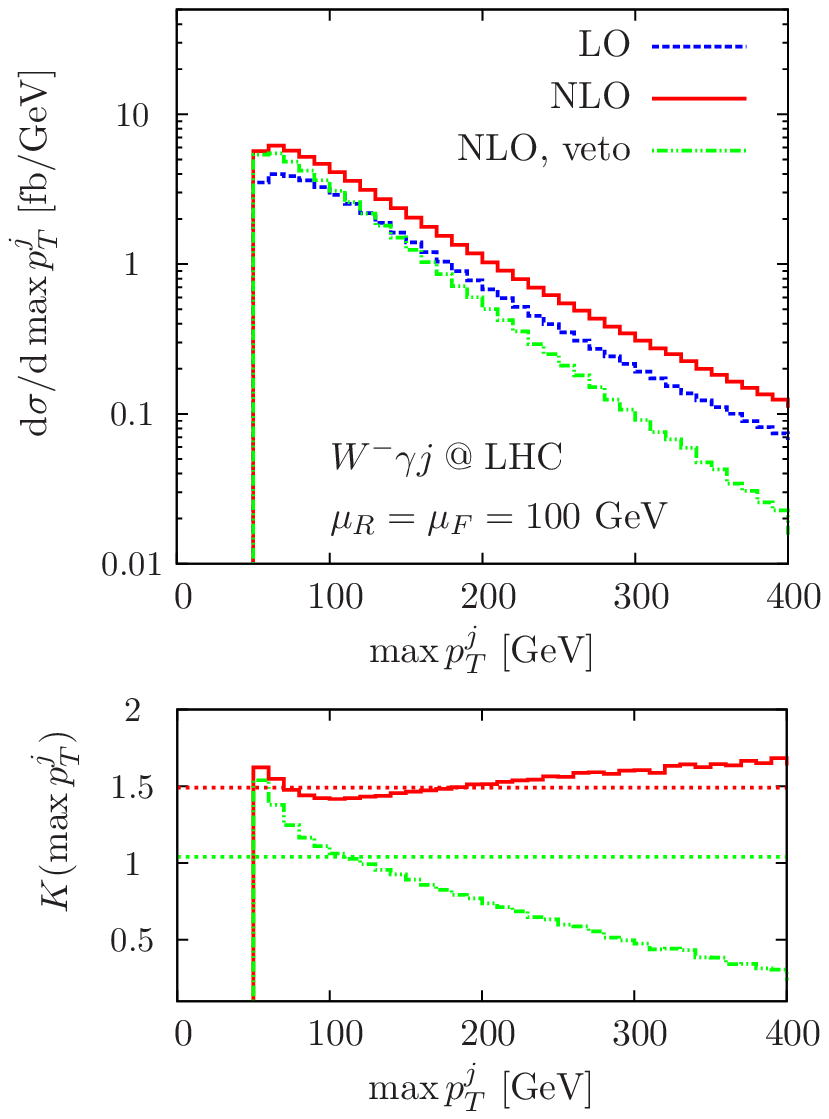, width=0.47\textwidth}
\caption{\small\label{fig:ptaptj} Transverse momentum of the $W$ and maximum jet transverse momentum at leading order (dashed) and next-to-leading order (solid). 
The lower panels display the differential $K$ factor, with the $K$ factors of the total inclusive and exclusive production plotted as dashed horizontal lines. 
The $W$ momentum is reconstructed from its decay products in the transverse plane.}
\end{center}
\end{figure}
\begin{figure}[t]
\begin{center}
\epsfig{file=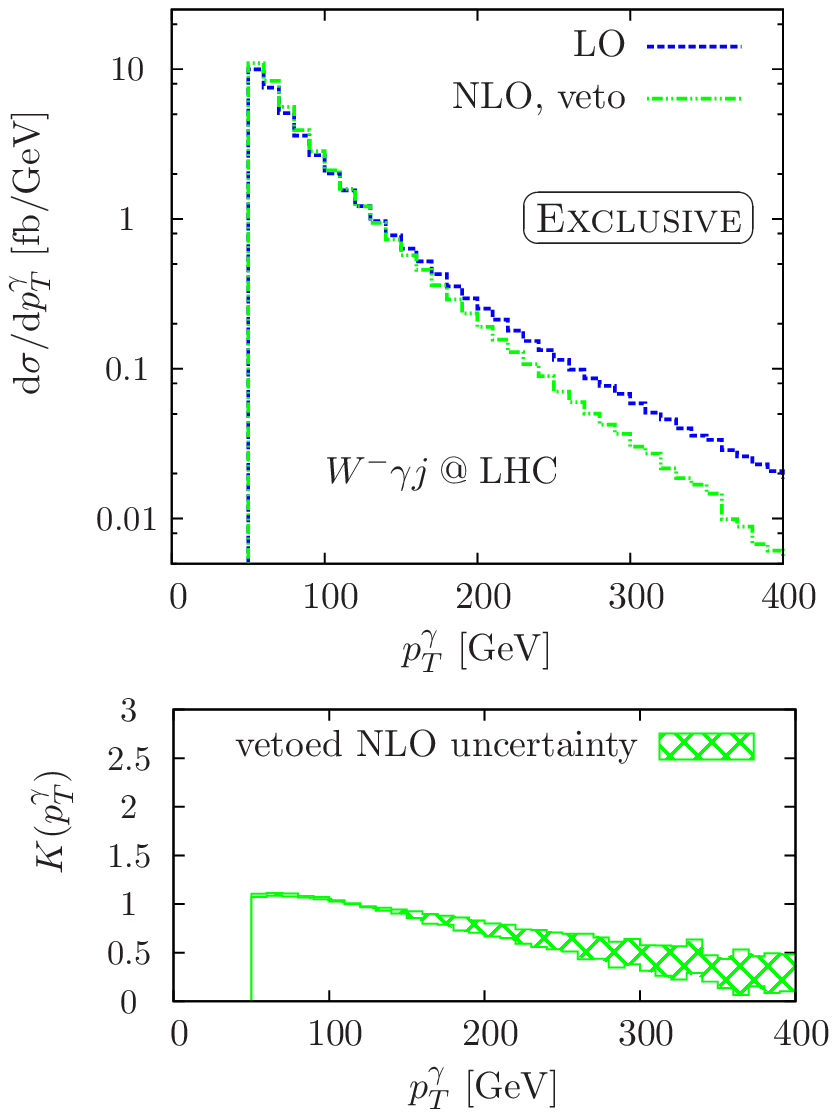, width=0.47\textwidth}
\hfill
\epsfig{file=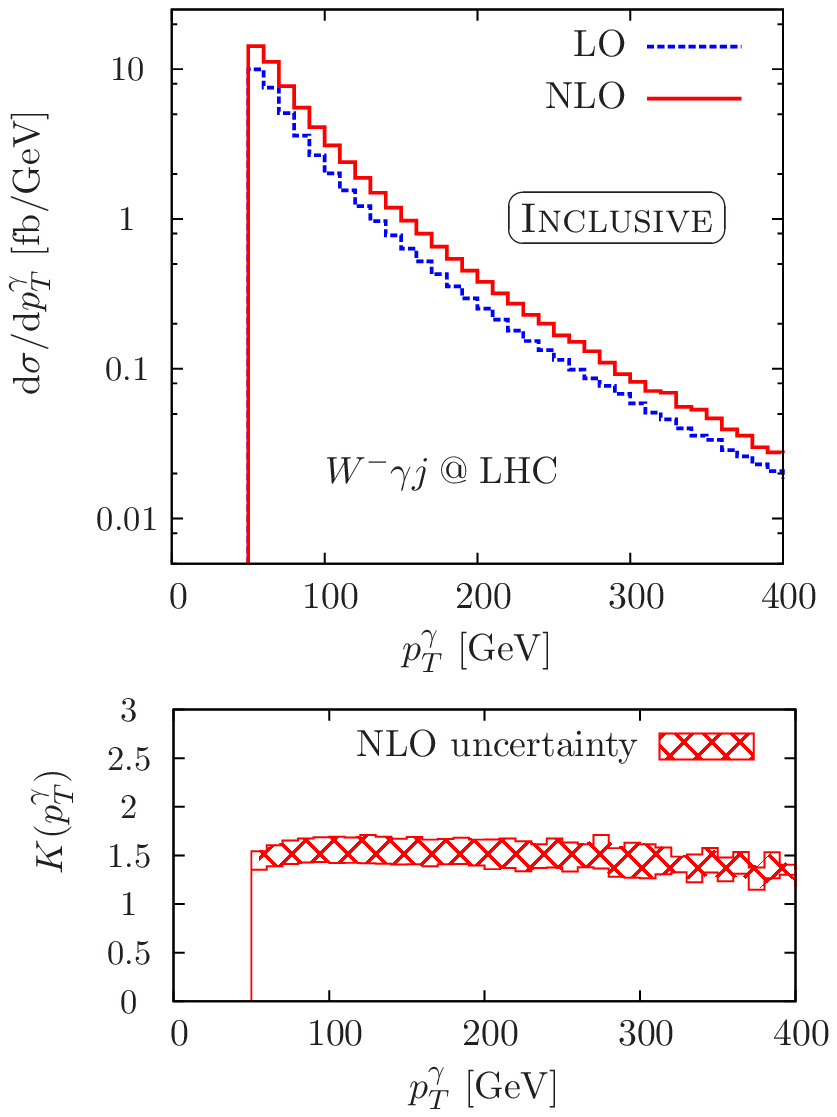, width=0.47\textwidth}
\caption{\small\label{fig:ptauncer} Transverse momentum of the photon for exclusive (left panel) and inclusive $W^-\gamma +{\rm{jet}}$ production (right panel). 
The uncertainty bands refer to fixed scale variations around $\mu_R=\mu_F=100\gev$ by a factor two in the NLO computation only.}
\end{center}
\end{figure}

%
We quote the inclusive cross sections at the LHC for the chosen cuts in Tab.~\ref{tab:kfactorswga}.
The differences comparing $W^+$ to $W^-$ is mainly due to the different parton distributions at the LHC.
The cross sections' scale dependencies, estimated by varying $\mu_F=\mu_R$ by a factor two around $100\gev$, is reduced from approximately $11\%$ 
to $7\%$ when including the NLO QCD corrections. As already pointed out, introducing a veto on the second resolved jet, leads to 
considerable stabilization of the integrated NLO cross section; the one jet-exclusive cross sections (Tab.~\ref{tab:kfactorswgaex}) exhibit scale variations
of $0.5\%$ for $W^-\gamma$+jet and $2\%$ for $W^+\gamma$+jet production, respectively. Naively, this suggests that vetoing the second hard jet amounts to a 
perturbative stabilization of the NLO cross section at a lower exclusive rate by effectively rejecting the $\mu_R$-dependence of the real emission dijet contributions 
to the hadronic NLO $W\gamma$+jet cross sections. 
However, given that extra jet radiation becomes important in the tails of the transverse momentum distributions,
Fig.~\ref{fig:ptaptj}, the veto introduces substantial uncertainties in this very phase space region. 
This can be inferred from Fig.~\ref{fig:ptauncer}, where we exemplarily examine the impact of the fixed-scale variation 
$\mu_F=\mu_R=50\gev,\dots,200\gev$ on the photon's transverse momentum for both inclusive and exclusive production.
For completeness we note that dynamical scale choices, such as $\mu_R=\mu_F=\max p_T^j$, result in quantitatively similar uncertainty bands.
While the distribution's uncertainty band's relative size is uniform over the entire range of the distribution for inclusive production, 
the jet veto stabilizes $\d\sigma_{\rm{excl}}^{\rm{NLO}}/\d p_T^\gamma$ exclusively in threshold region, which dominates the integrated  
exclusive rate. 
In fact, the distributions for $\mu=50\gev$ and $\mu=100\gev$, which are used to generate the uncertainty band in 
Fig.~\ref{fig:ptauncer}, intersect at $p_T^\gamma\approx 100\gev$ signaling an accidental cancellation of the renormalization and factorization 
scale dependence for the small total $K$ factors of the exclusive sample. The large uncertainty in the exclusive $p^\gamma_T$ distribution's tail 
then translates into an only mild overall scale dependence $\sigma_{\rm{excl}}^{\rm{NLO}}$. Similar conclusions have been drawn for 
$WZ$+jet production in Ref.~\cite{yettoapp}.
\begin{table}[!t]
\begin{center}
\begin{tabular}{c c c}
\hline
& 
$\sigma^{\rm{NLO}}_{\rm{excl}}$ [fb]& 
$K=\sigma^{\rm{NLO}}_{\rm{excl}}/\sigma^{\rm{LO}}$ \\
\hline
$W^-\gamma j$  & $429.2\pm 0.8$   &  $1.04$ \\
$W^+\gamma j$ & $495.1 \pm 1.0$   &  $1.06$ \\ 
\hline
\end{tabular}
\caption{\small\label{tab:kfactorswgaex} Exclusive next-to-leading order cross sections and total 
$K$ factors for the processes $pp\rightarrow e^+\nu_e\gamma j+X$ and 
$pp\rightarrow e^-\bar\nu_e\gamma j+X$ at the LHC for identified renormalization and factorization scales, 
$\mu_R=\mu_F=100~\rm{GeV}$. The cuts are chosen as described in the text.}
\end{center}
\end{table}

%
%

\begin{figure}[t]
\begin{center}
\epsfig{file=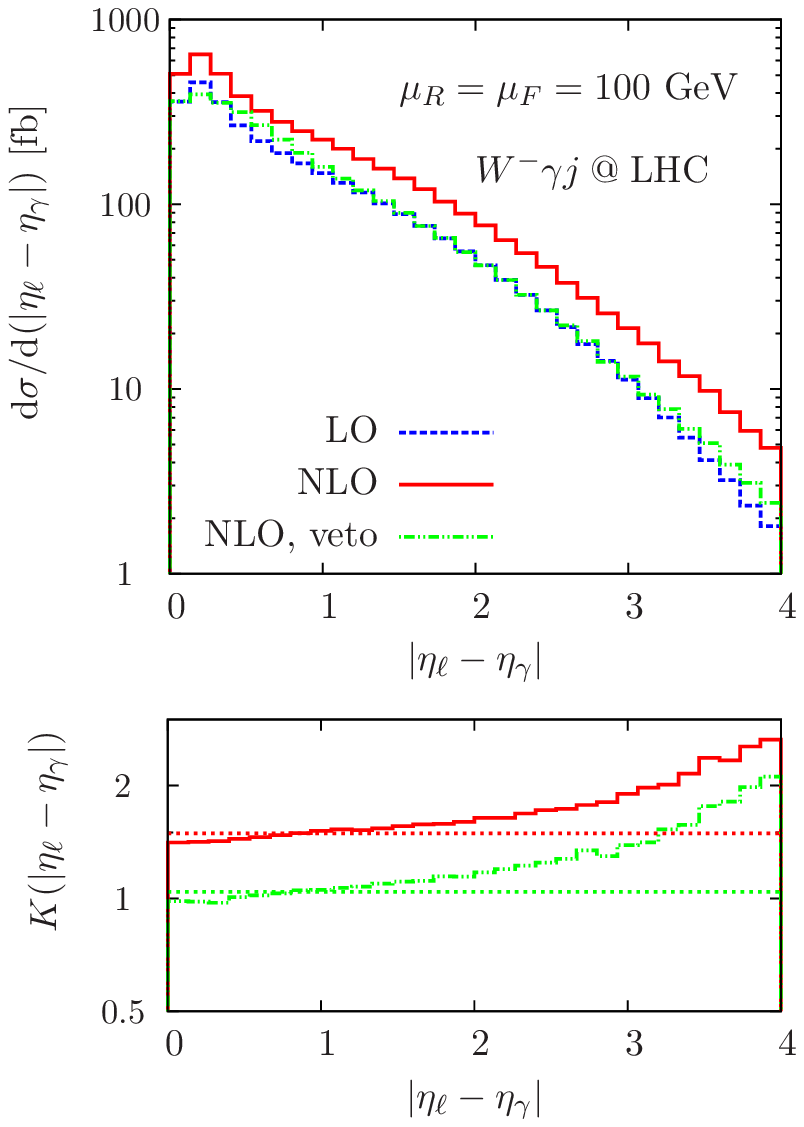, width=0.47\textwidth}
\hfill
\epsfig{file=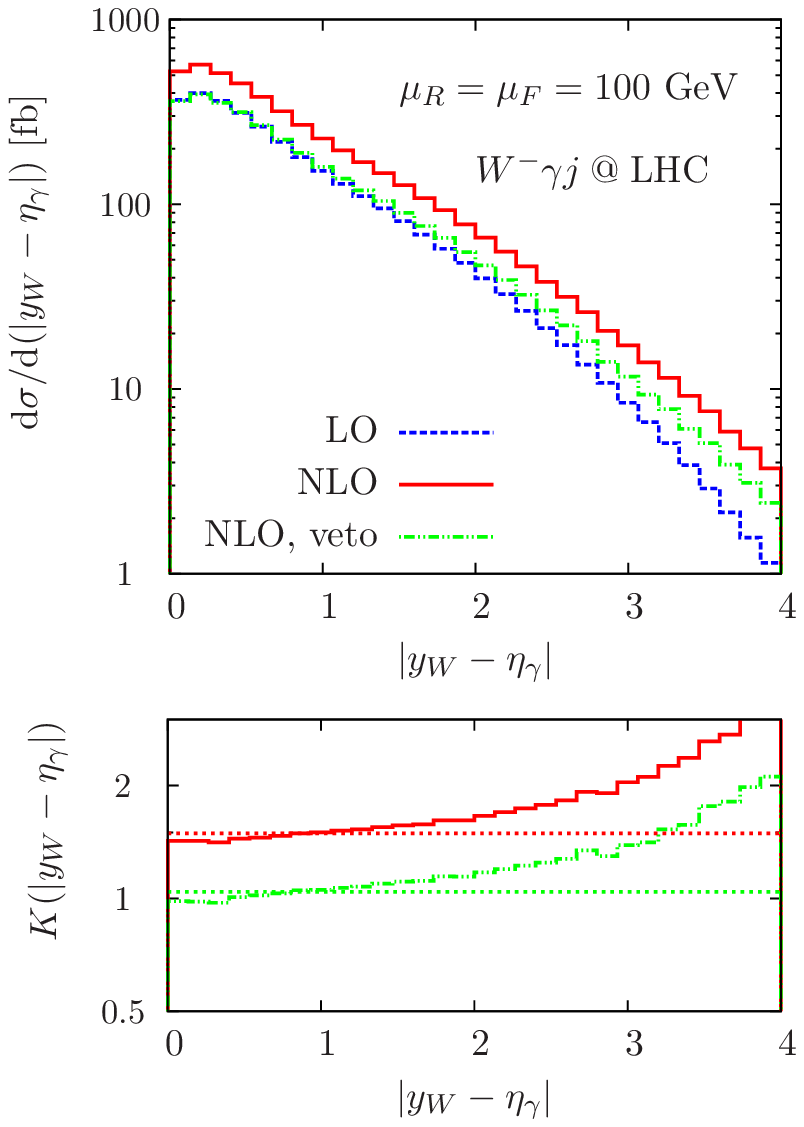, width=0.47\textwidth}
\caption{\small\label{fig:yl}  Lepton-photon and $W$-photon (pseudo)rapidity separation LO and NLO. The horizontal lines display
the $K$ factors of the total inclusive and exclusive production. The $W$ four-momentum
is defined from its decay products: $p^W_\mu=p^\ell_\mu +{\slashed{p}}_\mu$.}
\end{center}
\end{figure}

\begin{figure}[t]
\begin{center}
\epsfig{file=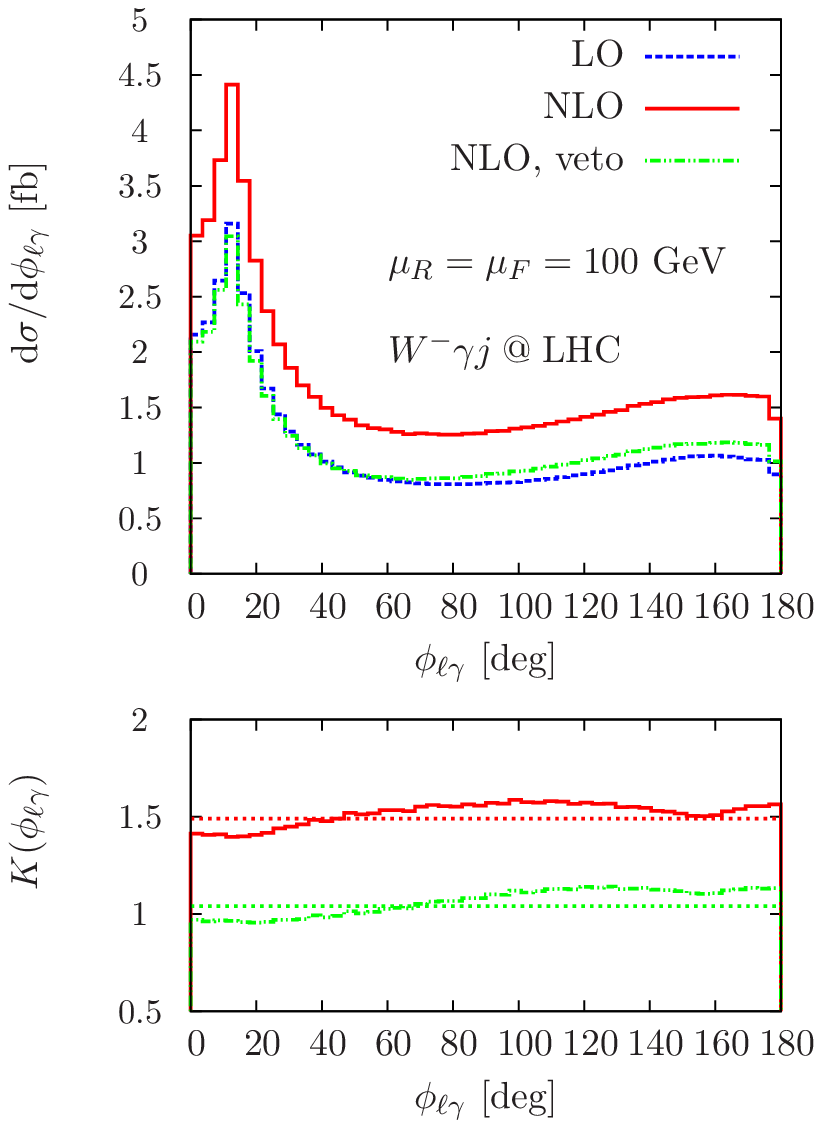, width=0.47\textwidth}
\hfill
\epsfig{file=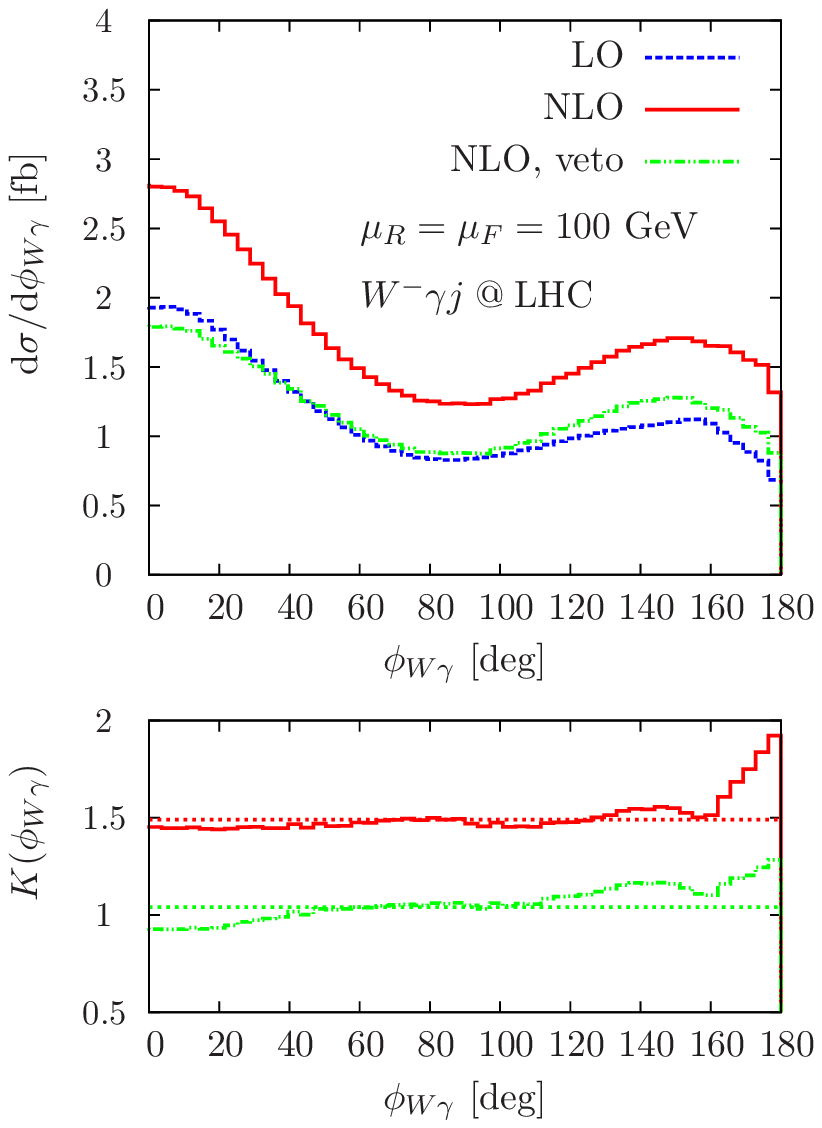, width=0.47\textwidth}
\caption{\small\label{fig:phis} Leading order and next-to-leading order distributions of the azimuthal angle between photon and lepton, and photon and $W$. 
The horizontal lines display the $K$ factors of the total inclusive and exclusive production.}
\end{center}
\end{figure}

This is yet another example of the well-known fact that 
total $K$ factors and total scale variations tend to be misleading when quantifying the impact of QCD quantum corrections to a given process. 
A better understanding of the QCD effects can be gained from differential $K$ factors of (IR-safe) observables $\cal{O}$,
%
%
%
\begin{figure}[t]
\begin{center}
\epsfig{file=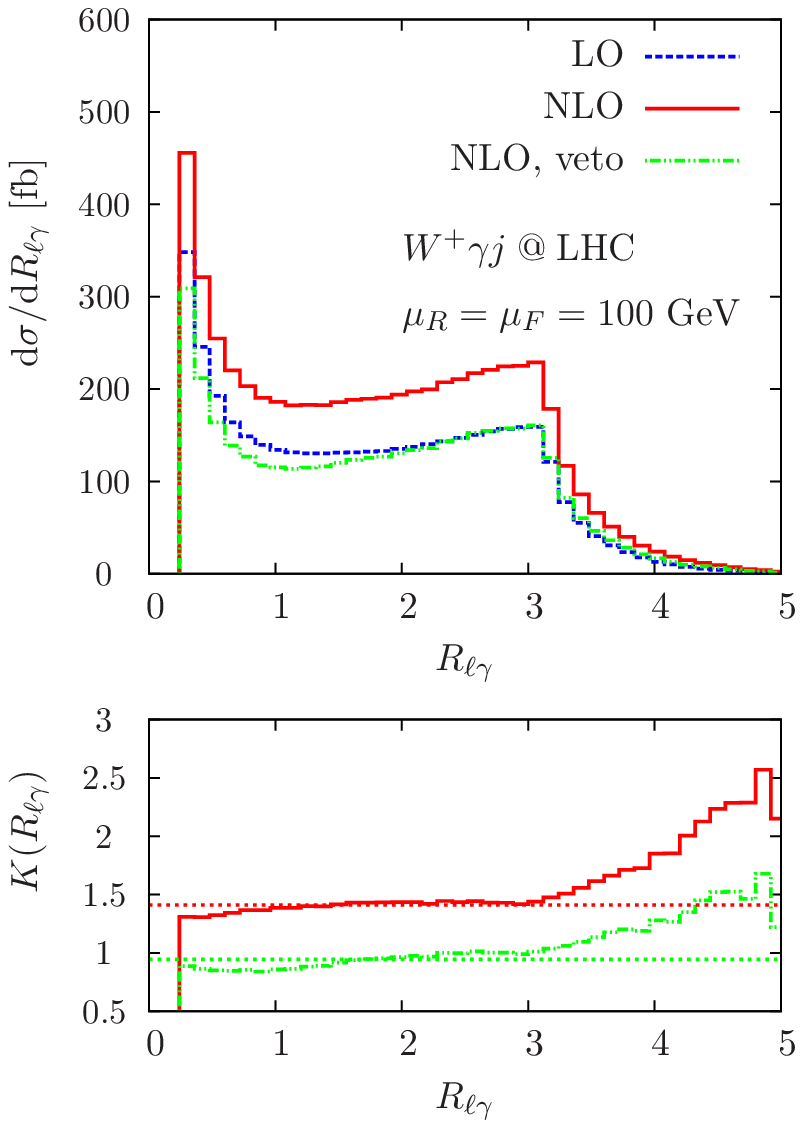, width=0.47\textwidth}
\hfill
\epsfig{file=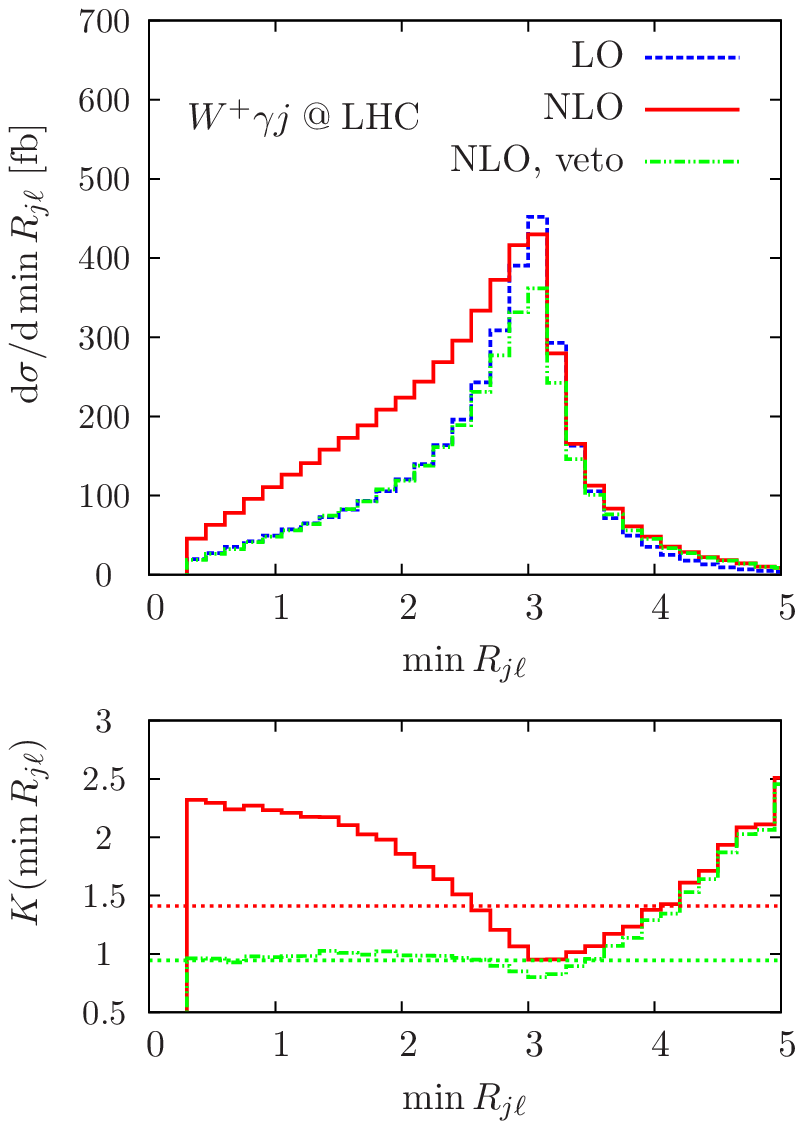, width=0.47\textwidth}
\caption{\small\label{fig:rlarjl} Lepton-photon separation and minimum jet-photon separation at LO and NLO. The horizontal lines display
the $K$ factors of the total inclusive and exclusive production.}
\end{center}
\end{figure}
\bee
\label{eq:diffkfac}
K({\cal{O}})=\ta{\sigma^{\rm{NLO}}}{\cal{O}}\bigg/\ta{\sigma^{\rm{LO}}}{\cal{O}} \,,
\eee
which encode the phase space dependence of the corrections, projected onto the respective observable. 
With unsuppressed extra jet radiation  at the LHC's large center-of-mass energy\footnote{The proton is 
probed at small momentum fractions $x\sim 0.02$ at LO, which results in $\sigma^{\rm{LO}}(W^\pm\gamma j)/\sigma^{\rm{LO}}(W^\pm\gamma jj)\sim 2$ 
for the chosen selection criteria. Note that this is qualitatively different from the situation encountered in NLO $W\gamma$ production.},
the distributions' shapes are highly altered when including NLO inclusive corrections. While the shapes of purely electroweak distributions,
i.e. distributions of observables that involve only photon, lepton and missing energy, survive to NLO QCD for the most part of the phase space, 
semi-hadronic observables get significantly modified with respect to their LO approximations due to additional jet radiation. 
Representatively, we show the azimuthal angle--pseudorapidity separation of lepton and photon and the minimal distance between jet and lepton in 
Fig.~\ref{fig:rlarjl}. We also plot  the azimuthal angle between lepton and photon and $W$ and photon in Fig.~\ref{fig:phis}. The purely leptonic observables 
also receive sizable modifications at the edges of the phase space, which are determined by the chosen cuts. 
If, e.g., the $W\gamma$ system recoils against additionally emitted partons at NLO, the $W$ and the photon are forced
to larger rapidity differences (Fig.~\ref{fig:yl}), which communicates to the azimuthal angle--pseudorapidity separation of lepton and photon at large values. 
Obviously this effect cannot be buffered by the additional jet veto. It is important to note that the pseudorapidity difference of the $W$ and the photon in Fig.~\ref{fig:yl} 
is experimentally not observable because the neutrino's longitudinal momentum explicitly enters the observable's definition. We 
show the distribution for comparison only. All other observables do depend only on the missing transverse momentum, 
which is experimentally reconstructed from the event's calorimeter entries \cite{Aad:2008zzm}.

%
%

%
%
%
%
The typical signatures of $W\gamma {\rm{+jet}}$ production at the LHC are dominated by configurations close to the
$p_T$-thresholds; the entire event is central with lepton and photon preferably emitted 
at small angular distances in the transverse plane. The collinear photon-lepton singularity is cut away by the requirement of Eq.~\gl{eq:lgammacut}.  
The photon is typically emitted collinear to the $W$ in the transverse plane. The jet recoils against the 
$W\gamma$ pair, and is back-to-back to the $W$ and the photon in the azimuthal angle distribution. 
For events with $W$ and $\gamma$ back-to-back (e.g. $ |\phi_{W\gamma}|\geq 150~{\rm{deg}}$, where the inclusive NLO cross sections 
drops by about $80\%$ to $\sigma^{\rm{NLO}}_{\rm{incl}}=131~{\rm{fb}}$) the jets tend to balance each other at small rapidity gaps of order one. These gaps
are due to the dominant $qg$- and $\bar q g$-induced partonic subprocesses of the dijet contribution. 
Due to the relatively large available phase space for additional jet emission,
the corrections are particularly large for this phase space region, $K=1.6$. 
This is in accordance with the large differential $K$ factor around $\phi_{W\gamma} \lesssim \pi$ in Fig.~\ref{fig:phis}. 
Note that this is also a part of the phase space which is more sensitive to the distinct photon isolation scales. In addition, with $W$ and photon back-to-back,
events with $ |\phi_{W\gamma}|\geq 150~{\rm{deg}}$ can be understood as ``genuine'' $pp\rig W\gamma+X$ events, subject to anomalous couplings' 
studies\footnote{These configurations give rise to large momentum transfers in the trilinear $WW\gamma$ coupling. The QCD corrections
are therefore important to understand the deviations that result from anomalous couplings. We will discuss this in more detail
in Sec.~\ref{sec:anomalous}.}.

Anomalous couplings generically modify the $p_T^\gamma$ distribution at large values, Eq.~\gl{eq:anomform}. 
It is therefore worth commenting on the impact of the NLO corrections onto the region of phase space characterized by very large $p_T^\gamma$ 
already at this point. Considering energetic events in the tails of the $p_T$ distributions (e.g. $p_T^\gamma >1\tev$) the picture is 
quite different from the situation we have described above. Jet emission is logarithmically enhanced in the dominant 
gluon-induced subprocesses $qg\rig W\gamma Q$ already at LO, which can easily be seen from the Altarelli-Parisi \cite{Altarelli:1977zs}
approximation of collinear emission $q\rig Q W$ as demonstrated in Ref.~\cite{Baur:1993ir}
\bee
\label{eq:collinearqw}
\d \sigma (qg\rig W\gamma Q) =\d\sigma (qg\rig q\gamma) {e^2\over 16\pi^2\sin^2\theta_w} \log^2 {p^{\gamma}_T\phantom{}^2\over m_W^2}\,,
\eee
for a diagonal CKM matrix.
The preferred situation is therefore a collinear $W$-jet pair that recoils against the hard photon.
This region of phase space receives sizable QCD corrections: 
The extra parton emission in these events has no preferred direction in the azimuthal angle and is kinematically unsuppressed.
The $p_T^W$ distribution receives sizable corrections for the same reason.
The uncertainties in this region of phase space are dominated by the dijet contribution, which is only determined to LO 
approximation in our calculation. Hence, our NLO correction does not improve the cross sections' stability in this extreme region of phase space. However, as the
anomalous couplings affect the $p_T^\gamma$ distribution at values much lower than $\Lambda\sim{\rm{TeV}}$, see Eq.~\gl{eq:anomform},
the NLO corrections give rise to perturbatively predictive deviations from the SM, see Sec.~\ref{sec:anomalous}.

%
%
%
\begin{figure}[t]
\begin{center}
\epsfig{file=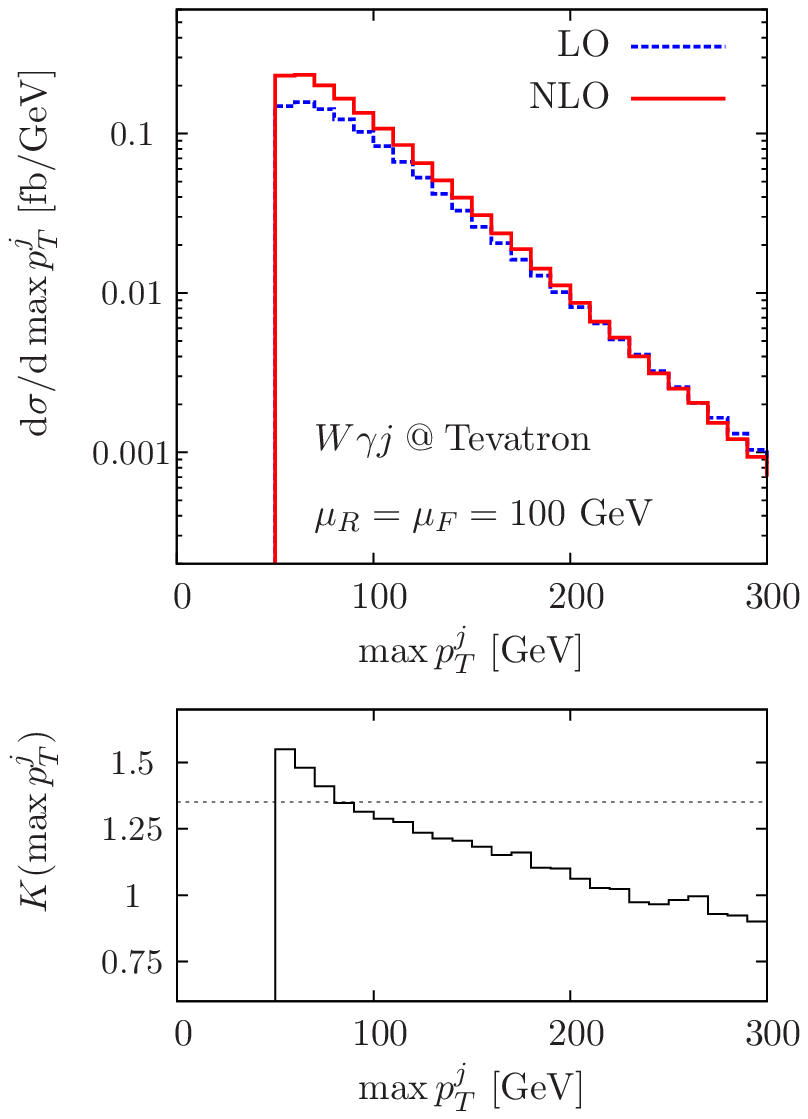, width=0.47\textwidth}
\hfill
\epsfig{file=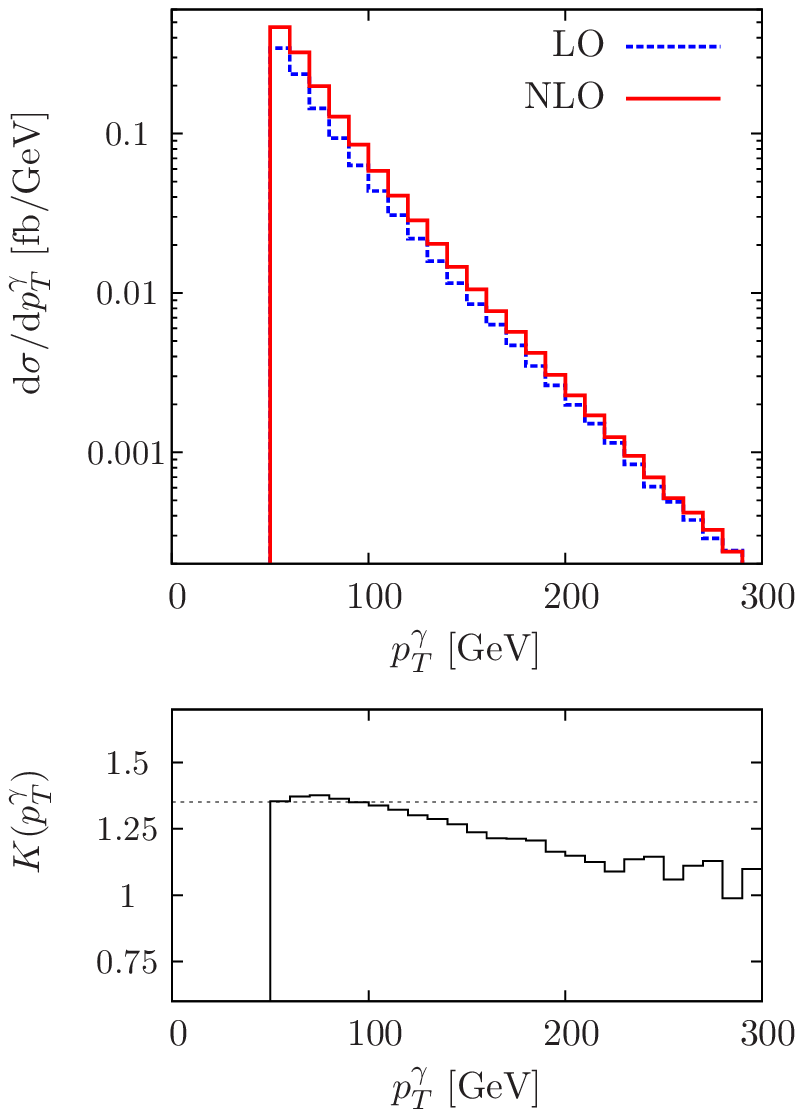, width=0.47\textwidth}
\caption{\small\label{fig:tevawgaptjptw} Leading order and next-to-leading order $\max(p_T^j)$ and $p_T^\gamma$ distributions at the Tevatron, 
including the respective differential $K$ factors. The horizontal line represents the $K$ factor of the total inclusive production.}
\end{center}
\end{figure}
%
%
%
\begin{figure}[t]
\begin{center}
\epsfig{file=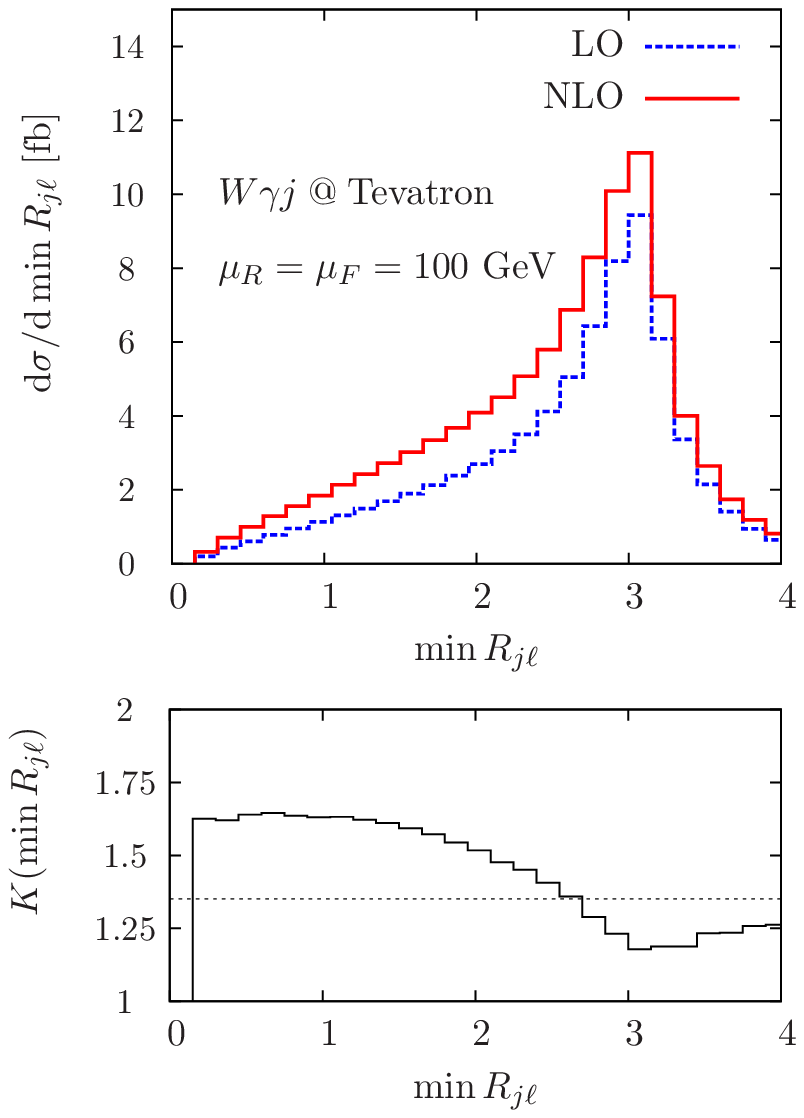, width=0.47\textwidth}
\hfill
\epsfig{file=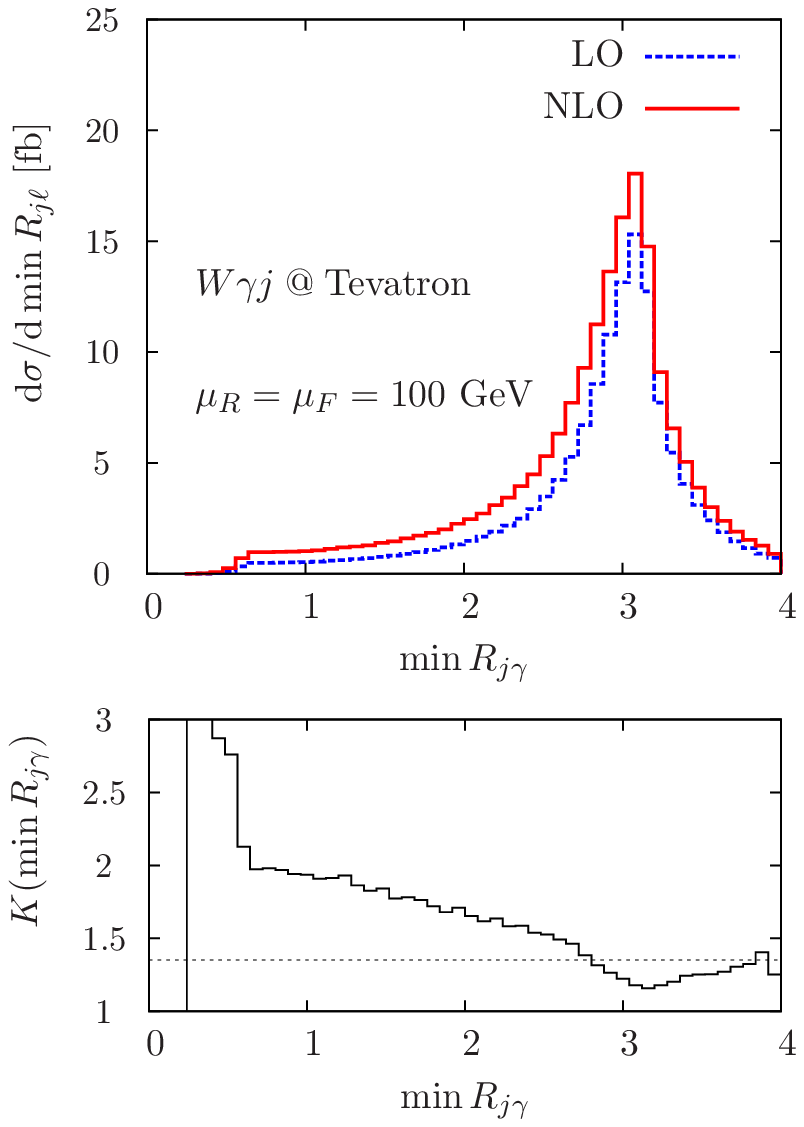, width=0.47\textwidth}
\caption{\small\label{fig:tevagarsep} Leading order and next-to-leading order $\min R_{j\ell}$ and $\min R_{j\gamma}$ 
distributions at the Tevatron, including the respective differential $K$ factors. The horizontal line gives the $K$ factor of the total inclusive production.}
\end{center}
\end{figure}
Vetoing the second resolved jet, Eq.~\gl{eq:wajveto}, removes most of the characteristics of additional jet radiation. The total exclusive cross sections are tabulated in 
Tab.~\ref{tab:kfactorswgaex}. 
Comparing the uncertainty band of the photon transverse momentum in Fig.~\ref{fig:ptauncer} and the 
respective phase space dependence of the QCD corrections in Fig.~\ref{fig:ptaptj}, we conclude that perturbative 
stability against variations of factorization and renormalization scales of dijet-vetoed $p p \rig \ell\nu\gamma j +X$ production 
shows up as a subtraction of a ${\cal{O}}(\alpha^3\alpha_s^2)$  leading-order contribution (the real emission dijet contribution) from a relatively 
stable inclusive NLO prediction. 
The vetoed contribution is kinematically well-accessible and unsuppressed by QCD dynamics.
The larger scale dependence of the vetoed distributions' tails compared to inclusive production remains as an echo. 
At larger values of $p_T^\gamma$, exclusive production does not yield a perturbatively reliable result, which is also indicated
by negative weights. 

\subsection{NLO QCD $W\gamma j$ production in the SM at the Tevatron}
\label{sec:tevares}
For Tevatron collisions we find a total cross section of
\bee
\label{eq:tevanlo}
\sigma^{\rm{NLO}} =(14.86\pm 0.03)~{\rm{fb}} \qquad (K=1.35)\,.
\eee
The proton and the antiproton are tested at $x\sim 0.2$ so that $q\bar Q$-induced subprocesses 
dominate the total hadronic cross section. 
The considerably lower center-of-mass energy compared to the LHC in combination with the cuts on the jet transverse 
momentum effectively introduces a jet veto, so that the Tevatron shapes resemble the NLO exclusive LHC distributions. 
The shapes of the transverse momentum distributions at large $p_T$  are overestimated by the LO
approximation, yielding differential $K$ factors of order 0.5 at NLO for the distributions' tails. Since additional jet radiation is kinematically 
suppressed compared to the LHC, the semi-hadronic observables typically receive smaller relative corrections around 
the $\sigma^{\rm{NLO}}/\sigma^{\rm{LO}}$ rescaled LO distributions. Yet, QCD-radiation effects are still sizable, and events 
tend to be re-distributed to smaller minimum separations of the hadronic jets with respect to the lepton and the photon, 
Fig.~\ref{fig:tevagarsep}.
\subsection{NLO QCD W${\bf{\gamma}}$+jet production with anomalous WW$\bf{\gamma}$ couplings at the LHC}
\label{sec:anomalous}
\begin{figure}[t]
\begin{center}
\parbox{7cm}{
\epsfig{file=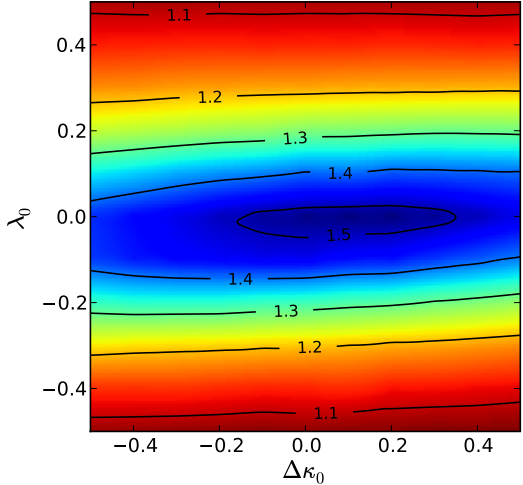, scale=0.6}
}
\hskip 1cm
\parbox{7cm}{
\caption{\small\label{fig:anok} Total $K$ factor contours for $pp\rig e^-\bar \nu_e \gamma j+X$ cross section at the LHC for anomalous input parameters $|\Delta\kappa_0|, |\lambda_0| \leq 0.5$
with dipole form factor $n=2$ and the cutoff scale $\Lambda=1\tev$.}
}
\end{center}
\end{figure}

We now include anomalous couplings to the NLO $W\gamma +\jet$ cross section predictions.
Most stringent bounds on anomalous $WW\gamma$ couplings are currently given by the combined 
analysis of LEP data of Ref.~\cite{Alcaraz:2006mx},
\begin{subequations}
\label{eq:anombounds}
\bee
1+ \Delta\kappa_0 = 0.984^{+0.042}_{-0.047}\,,\qquad
\lambda_0= - 0.016^{+0.021}_{-0.023}\,,
\eee
and recent fits at hadron colliders are from the Tevatron D${\slashed{0}}$ experiment Ref.~\cite{Abazov:2009hk}
\bee
1+ \Delta\kappa_0 = 1.07^{+0.16}_{-0.20}\qquad
\lambda_0 =- 0.0^{+0.05}_{-0.04} \,.
\eee  
\end{subequations}
Both bounds are at $68\%$ confidence level, extracted from data assuming $\Lambda=2\tev$ and dipole profiles. 
Note, that both experiments are consistent with the SM prediction $\Delta\kappa_0=\lambda_0=0$. Generically,
these bounds select a region in the parameter space where the QCD corrections are particularly important. This can
be inferred from a scan over a wide (and experimentally ruled-out) range of anomalous parameters in Fig.~\ref{fig:anok}. 
We choose cuts in resemblance to the selection criteria that are typically applied 
by the ATLAS collaboration to probe anomalous trilinear couplings, e.g., Ref.~\cite{Dobbs:2005ev}
%
%
%
\begin{figure}[t!]
\begin{center}
\epsfig{file=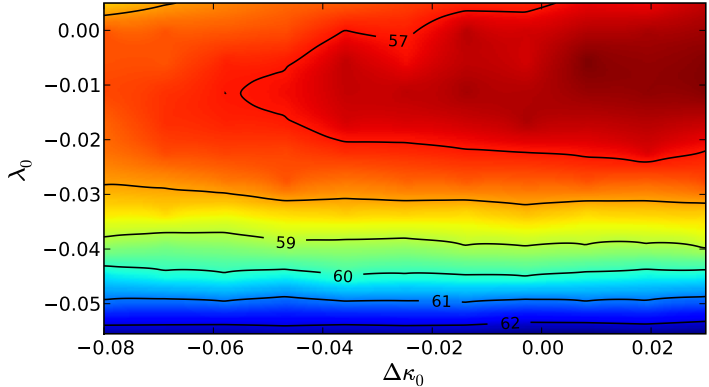,scale=0.6}
\caption{\small
\label{figcomb} Inclusive NLO QCD $pp\rig e^-\bar \nu_e \gamma j+X$ cross section contours at the LHC in fb after applying
the cut of Eq.~(3.15). We show the $W^-\gamma$+jet cross section for parameters $\kappa_0,\lambda_0$ that are consistent with the LEP bounds.
We choose dipole form factors $n=2$ and a cutoff scale $\Lambda=2\tev$ \cite{Dobbs:2005ev,Abazov:2009hk}.
}
\end{center}
\end{figure}
\bee
p_T^\gamma \geq 100\gev\,, \quad p_T^\ell,\slashed{p}_T\geq 25\gev\,, \quad R_{\ell\gamma}\geq 1.0 \,,
\eee
where $\slashed{p}_T$ denotes the missing transverse momentum. In addition, we choose inclusive hadronic jet cuts
\bee
p_T^j\geq 20\gev\,,\quad \delta_0=0.4\,, \quad R_{j\ell} \geq 0.2\,.
\eee 
\noindent 
These cuts yield a too low total rate at the Tevatron to be phenomenologically important. This can also be
inferred from comparing the $p_T^\gamma$ distributions at the LHC and the Tevatron for large values, where
the effects of anomalous couplings will be visible, Figs. \ref{fig:ptauncer} and \ref{fig:tevawgaptjptw}. We therefore
focus on anomalous couplings at the LHC.
The qualitative reason why the QCD corrections turn out large for parameter choices in the vicinity of the SM is easily 
uncovered by examining the corrections' $p_T^\gamma$ dependence.
From Fig.~\ref{fig:ptaptj} we infer\footnote{We slightly abuse the notation: $K(p_T^\gamma)$ is the differential
$K$ factor in the sense of Eq.~\gl{eq:diffkfac} and $K$ without parentheses refers to the total $K$ factor of exclusive
production.} that $K(p_T^\gamma)>K$ in the threshold region and $K(p_T^\gamma)<K$ in the tail of the distribution. 
Consequently, the region of phase space, where the anomalous couplings' impact is well-pronounced, i.e. $p_T^\gamma\lesssim \Lambda$, 
provides a smaller fraction to the NLO cross section for inclusive cuts compared to the LO approximation rescaled by the total $K$ factor of inclusive production. 
Additionally, at low transverse momenta, the distributions are dominated by SM physics due to small momentum transfers in Eq.~\gl{eq:anomform}, so that they 
are largely independent of $\Delta\kappa_0,\lambda_0$ in this particular phase space region. 
In total, not only a large fraction of the cross section, but also a large share of its increase compared to LO, is insensitive to the
underlying anomalous parameters for experimentally allowed values $\Delta\kappa_0,\lambda_0$. This is completely analogous 
to anomalous $WZ$+jet production \cite{wzano}. 
The NLO inclusive cross section is therefore less sensitive to the anomalous couplings than the LO cross section, and
$K=\sigma^{\rm{NLO}}/\sigma^{\rm{LO}}$ is large in regions, where the distributions are dominated by their low $p_T$ behavior: $K$ 
peaks around the SM, $\Delta\kappa=\lambda=0$, Fig.~\ref{fig:anok}. From Fig.~\ref{fig:ptauncer} and the related discussion, it is also 
apparent that, in addition to lost perturbative stability for large $p_T^\gamma$, the effects of anomalous couplings are suppressed in 
exclusive $W\gamma$+jet production.

\begin{figure}[t]
\begin{center}
\epsfig{file=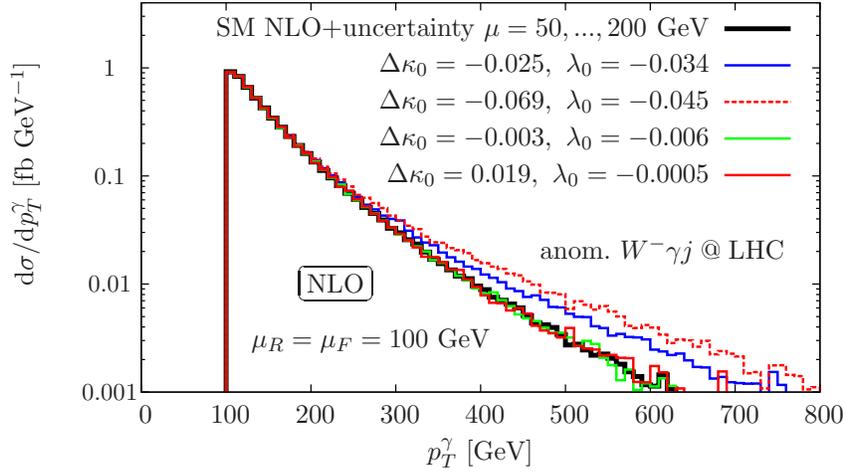,scale=0.92}
\caption{\small\label{figanompta} Inclusive NLO QCD distributions of the photon transverse momentum in anomalous
$pp\rig e^-\bar \nu_e \gamma j +X$ at the LHC for different parameters $\kappa_0,\lambda_0$ that are consistent with the LEP bounds.
We choose dipole form factors $n=2$ and a cutoff scale $\Lambda=2\tev$ \cite{Dobbs:2005ev,Abazov:2009hk}. The width of the SM curve represents the SM scale
uncertainty that results from varying $\mu_R=\mu_F$ around the central scale of 100 GeV by a factor two.}
\end{center}
\end{figure}

To increase the sensitivity to $\kappa_0$ and $\lambda_0$ in the experimentally allowed range, we additionally require
the $W$ and the photon to be back-to-back in the transverse plane, by imposing an azimuthal angle
\beeq
\label{eq:deltawgamma}
| \phi_{W\gamma}|\geq 150~{\rm{deg}}\,.
\eeeq
This cut effectively mimics ``genuine'' $W\gamma$ events with additional hadronic activity. Given the hard $p_T$
requirements, this selection criterion can be replaced by a cut on $| \phi_{\ell\gamma}|$ or $\Delta R_{\ell\gamma}$
without qualitatively changing the phenomenology (see also Figs. \ref{fig:yl} and \ref{fig:phis}).
The resulting variation of the integrated $W^-\gamma+\jet$ cross section for parameters $(\Delta\kappa_0,\lambda_0)$ in the range 
of Eq.~\gl{eq:anombounds} is of order 10\%, Fig.~\ref{figcomb}. Comparing this variation to the uncertainty inherent to the SM expectation
at the given order of perturbation theory, which, e.g., yields $\sigma\simeq 60.6~\rm{fb}$ for $\mu_R=\mu_F=50\gev$, we see that the 
cross sections' increase due to the anomalous couplings is compatible with the SM  NLO scale uncertainty, signaling a vanishing 
sensitivity of the total rate to $\Delta\kappa_0,\lambda_0$.

\begin{figure}[t]
\begin{center}
\epsfig{file=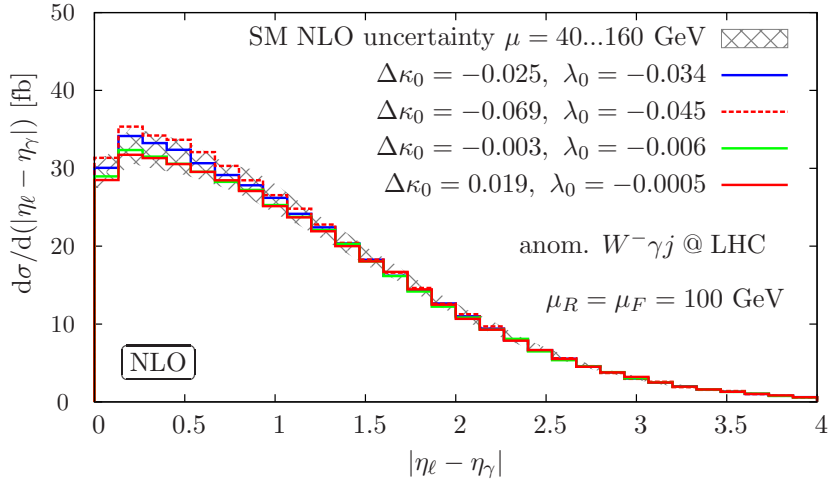,scale=0.92}
\caption{\small\label{figanomyla} Inclusive NLO QCD distributions of the photon lepton rapidity difference
in anomalous $pp\rig e^-\bar \nu_e \gamma j +X$ at the LHC. We show histograms with different 
parameters $\kappa_0,\lambda_0$ that are consistent with the LEP bounds.
We choose dipole form factors $n=2$ and a cutoff scale $\Lambda=2\tev$ \cite{Dobbs:2005ev,Abazov:2009hk}.}
\end{center}
\end{figure}

This, however, does not hold for differential distributions at large momentum transfers, 
e.g. for the $p_T^\gamma$ spectrum, which receives large anomalous couplings-induced modifications  
of the distribution's tail. The altered spectrum is well outside the SM-uncertainty band for larger values 
of $(\Delta\kappa_0,\lambda_0)$, with a particular sensitivity to $\lambda_0$. Remember that $\lambda_0$ dials the 
dimension six operator in Eq.~\gl{anovertex}, which is not present in the SM. The characteristic enhancement 
vanishes when the anomalous parameters approach their SM values, and the shape deviations 
become comparable to the distribution's uncertainty. The larger cross section at large $p_T^\gamma$ compared to the SM translates
into an increased cross section for the $W\gamma$ back-to-back configurations, which is also visible in
the pseudorapidity differences at small separation, Fig.~\ref{figanomyla}.
The anomalous couplings' impact on this distribution is qualitatively different from the 
QCD corrections, which exhibit $K(p_T^\gamma)<K$ for large $p_T^\gamma$. Therefore, the NLO cross section 
at small rapidity differences is smaller than the NLO-normalized LO distributions suggests, Fig.~\ref{fig:yl}.
Yet, the NLO uncertainty from integrating over the small $p_T$ configurations 
cover the anomalous couplings effect entirely, already by varying the scale within a small intervall,
as indicated in Fig.~\ref{figanomyla}.
Given the residual anomalous couplings-induced deviations of the $p_T^\gamma$ shape, a more inclusive 
measurement strategy that relies on fits to the inclusive $p_T^\gamma$ spectrum or on multivariate analysis of
the distributions can supplement traditional techniques and appears to be practicable. This is also motivated
by the overall theoretical uncertainty of order 10\% becoming comparable to the estimated experiments' systematics
at the reported order of perturbation theory.

\section{Summary and Conclusions}
\label{sec:summary}
In this paper, we have calculated the differential NLO QCD corrections to $W\gamma$ production in association with a hadronic jet,
including full leptonic decays and all off-shell effects of the $W$. We have given details on the calculation's strategy and have 
also discussed the effects of anomalous $WW\gamma$ couplings. The corrections are sizable and exhibit substantial phase space 
dependencies, and should be included in phenomenological analysis which employ these processes, either
as signal or as background.
In particular, we have found the exclusive production's perturbative stability to be
accidental --- additional jet vetoing does not amount to a experimental strategy which
is under good theoretical control at the given order of perturbation theory. 
Qualitatively identical results have been shown to hold for $WZ+\rm{jet}$ production in Ref. \cite{yettoapp}, where the dominant
QCD corrections are identical and only get tested at different scales. This
strongly indicates that the additional jet veto does not provide a meaningful procedure for the entire class of massive diboson plus 
jet cross sections beyond theoretical contemplation.

Even if theoretically less favored due to kinematical obstruction, inclusive
$W\gamma$+jet production (and hence inclusive $W\gamma$ production) 
exhibits potential sensitivity to anomalous couplings via shape deviations of the $p_T^\gamma$ 
distribution. Integrated cross sections for our inclusively  chosen leptonic cuts, however, entirely loose their sensitivity to 
modifications of the electroweak sector due to the low-$p_T$ QCD uncertainties. Providing the NLO corrections 
to $pp\rig \ell^\pm+\gamma+\slashed{p}_T+{\rm{jet}}+X$, we realistically
asses the impact of anomalous couplings on the characteristic $p_T^\gamma$ distribution in Fig.~\ref{figanompta}.
For larger anomalous couplings that are still compatible with the combined LEP measurements,
the distributions significantly deviate from their SM expectation and fall well outside the SM distribution's
uncertainty. Comparing to anomalous $WZ$+jet production, we find more sizable deviations in the distributions'
shapes in the allowed parameter range.
Whether the observed sensitivity can be carried over to the experiment represents
a challenging question, which is beyond the scope of this work. 
We leave a more thorough investigation of this direction to future work.

%

\subsubsection*{Acknowledgements}
We thank Dieter Zeppenfeld for collaboration during the initial stage of this work.
We also thank Steffen Schumann for {\sc Sherpa}-support. 
F.C acknowledges partial support by FEDER and Spanish MICINN under grant
FPA2008-02878. C.E. is supported in parts by the Karlsruhe Graduiertenkolleg 
``High Energy Particle and Particle Astrophysics''.
This research is partly funded by the Deutsche Forschungsgemeinschaft 
under SFB TR-9 ``Computergest\"utzte Theoretische Teilchenphysik'', and the 
Helmholtz alliance ``Physics at the Terascale''.


\end{document}